%Paper: astro-ph/9411116
%From: Rosemary Wyse <wyse@physics.Berkeley.EDU>
%Date: Wed, 30 Nov 1994 16:06:42 -0800 (PST)

%plain tex
\magnification=\magstep1
%\hoffset=2.5 true cm
%\voffset=3.5 true cm
%\vsize=22.5 true cm
\hsize=16.0  true cm
\parskip=0.2cm
\parindent=1cm
\raggedbottom

% \parfillskip=0pt right justifies the last line in a paragraph
\def\pp{\parshape 2 0truecm 16truecm 1truecm 15truecm}
%
%  \pp is a simple definition to define a paragraph shape in
%      which the first line is not indented, but subsequent lines are.
%      suitable for references and figure captions.
\def\apjref#1;#2;#3;#4 {\par\pp#1,  #2,  #3, #4 \par}
%
%   \apjref will construct a reference in post-1991
%           Ap.J style.  The first 3 arguments
%           must be delimited by semicolons,  and the last by a space.
%           all four arguments must be supplied, e.g.
%
% \apjref  Blow, Joe 1917;J.I.R.;1;991
\def\oldapjref#1;#2;#3;#4 {\par\pp#1, {\it #2}, {\bf #3}, #4. \par}
%
%   \oldapjref will construct a reference in pre-1991
%           Ap.J style.  The first 3 arguments
%           must be delimited by semicolons,  and the last by a space.
%           all four arguments must be supplied, e.g.
%
% \apjref  Blow, Joe, 1917;J.I.R.;1;991
\def\apj{ApJ}
\def\aap{A\&A}
\def\aaps{A\&AS}
\def\mnras{MNRAS}
\def\aj{AJ}
\def\araa{ARA\&A}

\def\apjs{ApJS}

%
% \header is a definition that gives a bf header, breaking the page
% 	if near the bottom
%
\def\subsection#1{\goodbreak\noindent\underbar{#1}}
\def\subsubsection#1{{\noindent \it #1}}
\def\ltsima{$\; \buildrel < \over \sim \;$}
\def\simlt{\lower.5ex\hbox{\ltsima}}
\def\gtsima{$\; \buildrel > \over \sim \;$}
\def\simgt{\lower.5ex\hbox{\gtsima}}
%
%
% VECTORS:
%{\b v} writes v in boldface italic.
%This contradicts tex82 which has \b meaning underline.
%\font\tfont=cmbi10
%\newfam\vecfam
%\def\vecfont{\fam\vecfam\tfont}
%\textfont\vecfam=\tfont \scriptfont\vecfam=\seveni
%\scriptscriptfont\vecfam=\fivei
%\def\b#1{\vecfont #1}
%{\bb v} writes v with an arrow above it.
%\def\bb#1{\skew{-2}\vec#1}

% MATH FUNCTIONS:
 %error function
 %hyperbolic sec
 %hyperbolic csc
 %arc hyperbolic sin
 %arc hyperbolic cos
 %arc hyperbolic tan
 %arc hyperbolic cot
 %arc hyperbolic sec
 %arc hyperbolic csc
 %arc cot
 %arc sec
 %arc csc
          %spherical harmonic
   %spherical harmonic primed
                               %real part
                               %imaginary part

% UNITS:

\def\kms{{\rm\,km\,s^{-1}}}

% FOOTNOTES:
\newcount\notenumber
\notenumber=1
%\note macro produces sequentially numbered footnotes at bottom of page
\def\note#1{\footnote{$^{\the\notenumber}$}{#1}\global\advance\notenumber by 1}

% EQUATION NUMBERING:
\newcount\eqnumber
\eqnumber=1
%\new macro produces sequentially numbered equations by writing \eqno\new)
%at end of displayed equations
\def\new{{\rm(\chaphead\the\eqnumber}\global\advance\eqnumber by 1}
%to refer to an equation which is 5 equations back, write "equation \ref5)"
\def\ref#1{\advance\eqnumber by -#1 (\chaphead\the\eqnumber
     \advance\eqnumber by #1 }
%\last macro is like \new except counter is not advanced. Useful for equations
%which are in parts a and b.
\def\last{\advance\eqnumber by -1 {\rm(\chaphead\the\eqnumber}\advance
     \eqnumber by 1}
\def\eq#1{\advance\eqnumber by -#1 equation (\chaphead\the\eqnumber
     \advance\eqnumber by #1}
%to name an equation, place command "\eqnam{\Poisson}{Poisson}" before
%equation, and
%thereafter "equation (\Poisson)" will generate the proper equation number.
\def\eqnam#1#2{\immediate\write1{\xdef\ #2{(\chaphead\the\eqnumber}}
    \xdef#1{(\chaphead\the\eqnumber}}

% FIGURE NUMBERING:
\newcount\fignumber
\fignumber=1
%\nfig macro assigns number to a figure
\def\nfig{\the\fignumber\ \global\advance\fignumber by 1}
%\nfiga permits a,b,c etc. to be added to figure number
\def\nfiga#1{\the\fignumber{#1}\global\advance\fignumber by 1}
\def\rfig#1{\advance\fignumber by -#1 \the\fignumber \advance\fignumber by #1}
\def\fignam#1#2{\immediate\write1{\xdef\
#2{\the\fignumber}}\xdef#1{\the\fignumber}}

% ABSTRACTS:
\newbox\abstr
\def\abstract#1{\setbox\abstr=\vbox{\hsize 5.0truein{\par\noindent#1}}
    \centerline{ABSTRACT} \vskip12pt \hbox to \hsize{\hfill\box\abstr\hfill}}

% MISCELLANEOUS:
% angles in degrees

%\degg produces degree symbol so that 3\sec5 produces 3.`5 with the degree
%symbol and the period aligned.

%\sec produces arcsec symbol so that 3\sec5 produces 3."5 with the second
%symbol and the period aligned.

%\s produces tilde in mathmode or horizontal mode.
\def\s{\ifmmode \widetilde \else \~\fi}
\def\={\overline}

\def\spose#1{\hbox to 0pt{#1\hss}}

\def\etal{{\it et al.\ }}
\def\cf{{\it cf.\ }}
\def\eg{{\it e.g.\ }}
\def\ie{{\it i.e.\ }}
%\lta and \gta produce > and < signs with twiddle underneath
\def\lta{\mathrel{\spose{\lower 3pt\hbox{$\mathchar"218$}}
     \raise 2.0pt\hbox{$\mathchar"13C$}}}
\def\gta{\mathrel{\spose{\lower 3pt\hbox{$\mathchar"218$}}
     \raise 2.0pt\hbox{$\mathchar"13E$}}}
%\Dt and \dt put Newton's notation dots above upper and lower case chars
\def\Dt{\spose{\raise 1.5ex\hbox{\hskip3pt$\mathchar"201$}}}	% upper case
\def\dt{\spose{\raise 1.0ex\hbox{\hskip2pt$\mathchar"201$}}}	% lower case

% TSS ADDITIONS:

\def\=={\equiv}

\def\dotsfill{\leaders\hbox to 1em{\hss.\hss}\hfill}

% EXTRA FONTS
%\font\sanserif=amss10
%\font\smallcap=amcsc10

%	This file contains definitions etc. to force Plain TeX to use the
%	old style American Modern fonts rather than the newer Computer
%	Modern fonts which are standard with the latest release of TeX.
%	To use this, use \input oldfonts somewhere near the top of your
%	TeX file.

%\font\tenrm=amr10
%\font\teni=ammi10
%\font\tensy=amsy10
%\font\tenbf=ambx10
%\font\tensl=amsl10
%\font\tentt=amtt10
%\font\tenti=amti10
%\font\tenex=amex10
%\font\sevenrm=amr7 \font\seveni=ammi7 \font\sevensy=amsy7 \font\sevenbf=ambx7
%\font\fiverm=amr5 \font\fivei=ammi5 \font\fivesy=amsy5 \font\fivebf=ambx5

%\textfont0=\tenrm \scriptfont0=\sevenrm \scriptscriptfont0=\fiverm
%\def\rm{\fam0 \tenrm}

%\textfont1=\teni \scriptfont1=\seveni \scriptscriptfont1=\fivei
%\def\mit{\fam1 } \def\oldstyle{\fam1 \teni}

%\textfont2=\tensy \scriptfont2=\sevensy \scriptscriptfont2=\fivesy
%\def\cal{\fam2 }

%\textfont3=\tenex \scriptfont3=\tenex \scriptscriptfont3=\tenex

%\newfam\itfam \def\it{\fam\itfam\tenit} \textfont\itfam=\tenit
%\newfam\slfam \def\sl{\fam\slfam\tensl} \textfont\slfam=\tensl
%\newfam\bffam \def\bf{\fam\bffam\tenbf} \textfont\bffam=\tenbf
%   \scriptfont\bffam=\sevenbf  \scriptscriptfont\bffam=\tenbf
%\newfam\ttfam \def\tt{\fam\ttfam\tentt} \textfont\ttfam=\tentt

\rm		% preselect roman fonts
% American date ordering convention MD,Y:
\def\Atoday{\ifcase\month\or
  January\or February\or March\or April\or May\or June\or
  July\or August\or September\or October\or November\or December\fi
  \space\number\day, \number\year}
% European date ordering convention DMY:
\def\Etoday{\number\day\space\ifcase\month\or
  January\or February\or March\or April\or May\or June\or
  July\or August\or September\or October\or November\or December\fi
  \space\number\year}
      %default to European date ordering
%                         override with \def\today{\Atoday}.

 \def\aap{A\&A} \def\aaps{A\&AS} \def\kms{{\rm km s^{-1}}}
\baselineskip=12pt
 \centerline {{\bf A DETERMINATION OF}}
 \centerline {{\bf THE THICK DISK CHEMICAL ABUNDANCE
 DISTRIBUTION:}}
 \centerline {\bf {IMPLICATIONS FOR GALAXY  EVOLUTION} }
 \bigskip

 \centerline {GERARD GILMORE}
 \centerline {Institute of Astronomy}
 \centerline {Madingley Road, Cambridge CB3 0HA}
 \centerline {England, UK}
 \bigskip
 \centerline {ROSEMARY F.G.~WYSE}
 \centerline {Center for  Particle Astrophysics }
 \centerline {301 Le Conte Hall}
 \centerline {University of California}
 \centerline {Berkeley, CA 94720}
 \centerline {USA}
\smallskip
 \centerline {and}
\smallskip
 \centerline {Department of Physics and Astronomy\footnote{$^1$}{Permanent
 Address}}
 \centerline {The Johns Hopkins University}
 \centerline {Baltimore, MD 21218}
 \centerline {USA}
 \bigskip
 \centerline { J.~BRYN JONES}
 \centerline {Department of Physics and Astronomy}
 \centerline {University of Wales, College of Cardiff}
 \centerline {Cardiff, CF4 3TH}
 \centerline {Wales, UK}
 \bigskip
 \vfill\eject
 \centerline {ABSTRACT}

 We present a determination of the thick disk iron abundance
 distribution obtained from an {\it in situ\/} sample of F/G stars.
 These stars are faint, $15 \simlt V \simlt 18$, selected on the basis
 of color, being a subset of the larger survey of Gilmore and Wyse
 designed to determine the properties of the stellar populations
 several kiloparsecs from the Sun.  The fields studied in the present
 paper probe the iron abundance distribution of the stellar populations
 of the Galaxy at 500--3000pc above the plane, at the solar
 Galactocentric distance.  The derived chemical abundance distributions
 are consistent with no metallicity gradients in the thick disk over
 this range of vertical distance, and with an iron abundance
 distribution for the thick disk that has a peak  at
 $-0.7$dex. The lack of a vertical gradient argues against slow,
 dissipational settling as a mechanism for the formation of the
 thick disk.
 The photometric and metallicity data support a turn-off of the
 thick disk that is comparable in age to the metal-rich
 globular clusters, or $\simgt$12Gyr and are
 consistent with a  spread to older ages.

 \vfill\eject
 \bigskip
 \centerline  {1. INTRODUCTION}

 The metallicity distribution of complete samples of long-lived stars
 has long been recognised as providing unique constraints on the early
 stages of chemical evolution of the Galaxy.  The main sequence
 lifetime of F/G dwarf stars is greater than the age of the Galaxy and
 hence the chemical-abundance distribution function of such stars
 provides an integrated record of the chemical-enrichment  history
 without the need for model-dependent corrections for dead stars
 (\cf\ van den Bergh 1962; Schmidt 1963; Tinsley 1980).
 Pioneering studies focussed on the only reasonably-complete sample
 available, which is that for stars in the immediate solar
 neighborhood; in effect stars within about 30pc of the Sun.  These
 samples have been sufficiently small that reliable study of any
 stellar population whose kinematics are such that member stars spend
 only a small fraction of an orbit in the solar neighborhood has
 necessarily been difficult. This is potentially a serious restriction,
 as such stars might in principle be a major contributor to the stellar
 population in a valid, representative volume of the Galaxy.
 In addition, intrinsically-rare stellar populations may be missed entirely.

   The present paper extends previous work by analysing an {\it in
 situ\/} sample of F/G dwarfs with spectroscopically-determined iron
 abundances, at distances up to 5kpc from the Sun.  A companion paper
 presents a determination of the solar neighborhood metallicity
 distribution derived from new, high-precision intermediate-band
 photometry.  The combination of these data sets provides a composite
 distribution function which is the most reliable presently-available
 description
 of the integrated distribution of chemical abundances in a column
 through the local Milky Way.

 \bigskip
 \centerline{2. CHOICE OF STELLAR TRACERS}

 The ideal tracer of Galactic Structure is one which is selected
 without any biases, does not suffer fom stellar age-dependent selection
 effects,
 is representative of the underlying
 populations, and is easily observable.
 Historically, the need for easy observation restricted studies to the
 immediate solar neighborhood. The primary limitation of the
 nearby star sample is its small size.  This inevitably means that
 stars which are either intrinsically rare -- such as halo population
 subdwarfs -- and stars which are common but whose spatial distribution
 is such that their local volume density is small -- such as thick disk
 stars -- are poorly represented.  Most recent and current efforts to
 extend present local volume-limited samples to include minority
 populations have, for practical observational reasons, utilized
 kinematically-selected samples defined in the solar neighborhood,
 following the pioneering work of Eggen, Lynden-Bell and Sandage
 (1962).  Subsequent correction for the kinematic biases inherent in
 these samples requires careful modelling (\cf\ Norris and Ryan 1991;
 Ryan and Norris 1993; Aguilar \etal\ 1994).  An {\it in situ\/}
 sample, truly representative of the dominant stellar population far
 from the Sun, circumvents these large,
 model-dependent corrections.

 Several surveys of tracer stars which can be observed {\it in situ}
 are available.  Intrinsically luminous tracers are {\it a priori\/}
 favored in terms of telescope time, but the likely candidates have
 other characteristics that diminish their suitability: RR Lyrae stars
 have  intrinsic age and metallicity biases in that only stars of a given range
 in metallicity and age exist in this evolutionary stage; the
 accessible globular clusters are few in number; bluer horizontal
 branch stars are also rare, and their color distribution depends on
 chemical abundance and on the unidentified
 `second parameter(s)'. K-giants are the most representative {\it
 evolved\/} tracers of the spheroid, and have been used extensively.
 However, one must first identify giant stars from among the
 substantially larger number of foreground K dwarfs with similar
 apparent magnitudes and colors.

 A desirable solution to these limitations, which has become feasible
 with current multi-object spectroscopic systems and large-scale
 photometric surveys, is to identify and study F/G dwarfs to
 significant distances from the Sun.  This is the solution which we
 have adopted.
 Chemical abundances for these stars provide the integrated record of the
 star formation and enrichment history during
 the early stages of Galaxy formation, analogous to the local G-dwarf
 distribution.

 The photometric catalogs from which the present samples are
 selected  are
 derived from photographic plates from the UK Schmidt
 Telescope, and from the Las Campanas Observatory 2.5m telescope.
 These plates were scanned with the COSMOS and APM measuring
 machines, and calibrated using what have since become standard
 procedures.
 We have defined area-complete samples in several fields
 chosen specifically to optimise a Galactic structure analysis; those
 discussed here are the South Galactic Pole and UK Schmidt Field 117,
 at $(\ell,b) = (270, -45)$.  These two lines-of-sight are such that
 for the photometric definition as given below, we
 select stars at the same Galactocentric distance projected onto the
 plane, but at several kiloparsecs from the plane. The  fields
 were also selected to have low reddening from the Burstein and
 Heiles (1982) HI maps.

   The present samples were
 selected to have colors in the ranges $ 0.2 \simlt{\rm B-V }\simlt 1.0$,
 and $15 \simlt {\rm V }\simlt 18$.  The color range is designed to be wide
 enough to
  ensure that neither young metal-poor stars, which may
 be bluer than the dominant old metal-poor turnoff near B$-$V=0.4, nor very
 old metal-rich stars, whose turnoff color may be much redder than the
 turnoff of a metal-rich globular cluster, near B$-$V=0.5, are excluded.
 Standard star-count models indicate that the color-magnitude range
 isolated in the present survey should contain predominately thick-disk
 stars, with these making up at least 50\% of the total sample. Thin-disk
 stars should make up some 20\% of the total, with these being
 found almost exclusively at the reddest colors, while halo stars make
 up the remainder. (This prediction of course is the basis for our
 photometric and kinematic survey with this selection function.)
 These numbers are  not accurately known until an adequate
 sample of kinematic and chemical abundance data is available. Indications to
 date however, based on chemical abundance, kinematic and astrometric data,
 are that these models over-predict the actual number of halo stars by
 a factor of order five (\eg\ Cayrel \etal\ 1991a,b;
  Friel 1987;  Perrin \etal\ 1994).
 With that correction, one expects the thick disk
 to make up perhaps 70\% of the present sample, with the remainder
 being predominately old disk stars near the red edge of the sample
 selection.

 Calibration of photographic wide-field photometry is a complex subject;
 the methodology is described by Gilmore (1984),
 while specific examples of the techniques
 in practice, including application  to  the South Galactic
 Pole field discussed
 here, are  described in detail in Reid and Gilmore (1982),
 Gilmore and Reid (1983), Gilmore, Reid and Hewett (1985),
 Gilmore and Wyse (1985) and
 Kuijken and Gilmore (1989). The photometric standards upon
 which the calibration is based are described in those references, and in
 Stobie, Gilmore and Reid (1985) and Stobie, Sagar and
 Gilmore (1985).
 A detailed  presentation
 of the photometry in this survey, including that in several other
 fields in which abundance
 data are not yet available, will be presented elsewhere. For present
 purposes it is sufficient to note that the external calibration,
  \ie\   a
 comparison with standard stars to define the Pogson magnitude scale, has
 been checked from independent  CCD data by Gilmore and Wyse (1985). They
 showed that the zero point and magnitude scale of at least the
 SGP  dataset were well established, with error in scale and zero
 point being less than a few percent. That is, the data are accurate. We
 now consider precision.

 Determination of the internal photometric error,
 \ie\   the scatter in photometry, is extremely difficult to
 quantify reliably. This
 has been discussed for some of the data of relevance here by Gilmore, Reid
 and Hewett (1985) and by Reid and Gilmore (1982). A more detailed
 discussion, based on comparison of the independent Las Campanas and UK
 Schmidt data for the same fields, will be presented with the full
 photometric survey results. These studies are in agreement  that
  the South Galactic Pole data are characterised by a random
 photometric error  of 0.$^m$05 in B$-$V color, and 0.$^m$07 in V$-$I
 color, for V brighter than roughly V=17.5. At fainter magnitudes a noisy
 tail to the error distribution is evident. Since the precise location of
 the magnitude limit beneath which the photometry is well defined by a
 single Gaussian measuring error, without an over-populated tail to the
 error distribution, is very important for present  (chemical
 abundances) purposes; we consider
 this in more detail in the next section. In the second field,
 F117, the data
 indicate that the error distributions in both magnitude and color are
 well described by a single Gaussian with dispersion in color of
 0.$^m$05 in B$-$V and 0.$^m$07 in V$-$I, to at least V=18
 (Gilmore, Reid and Hewett 1985), or beyond the limit of
 the data discussed in this paper.

  The major telescope used for the spectroscopy was the 4m Anglo-Australian
 Telescope, first with the fibre plugboard system (FOCAP),
 then with the automated positioning fibre system (AUTOFIB), with the
 IPCS as detector.  Several thousand independent spectra of program
 stars, distributed in the various fields have been obtained.  These
 spectra are typically of $\sim$1.5\AA\ resolution, covering the wavelength
 range 4000-5000\AA, with a signal to noise ratio of around 10 for
 program stars.  The primary motivation for these spectra is to provide
 radial velocities accurate to about 10 $\kms$; the kinematics derived
 from the radial velocity distributions are discussed in separate
 papers (Gilmore and Wyse, in preparation)  as is the analysis of the
 photometry in all the fields.
  Although the spectra were not intended to be useful for
 abundance derivations, it proved possible to derive reliable
 abundances from a subset of the main sample, namely
 the relatively high signal-to-noise spectra of the
 cooler stars, using a new method described in detail elsewhere (Jones,
 Gilmore and Wyse 1995, hereafter JGW) and discussed briefly below.
 Here we focus on the chemical
 abundance distributions derived from those data.

 The selection of stars for spectroscopic
 observation is designed to provide clearly-defined samples limited by area
 and magnitude. A more complex selection function has arisen in the
 subsample for which we are able to derive chemical
 abundances. The most important
 additional parameter in determination of abundances is signal-noise ratio
 of the spectrum, with a color-dependence of the useful limiting value, as
 detailed in the next section. The effect of this limit is that
 observations during periods of unusually good and unusually bad seeing and
 clouds are favoured. Good seeing clearly captures more photons, while in
 very bad seeing and clouds brighter stars and long exposures were
 selected, to obtain some data. The effect of poor conditions is clearly
 seen in the data for F117, where a few stars much brighter than the
 majority have been observed successfully. In all cases, the selection of
 which stars to observe was made from the available area-
 and color-complete catalogs without systematic color bias. The requirement
that
 signal-noise be higher for hotter stars does, however, lead to a systematic
 bias, which we discuss and correct in detail below.

 \bigskip
 \centerline{3. THE IRON ABUNDANCE  DISTRIBUTION OF F/G DWARFS }
 \centerline {AT SEVERAL KILOPARSECS FROM THE PLANE}
 \smallskip
 \centerline  {\it 3.1 The Iron Abundance Indicator}

 The spectra of the F/G stars are of moderate resolution and
 signal-to-noise, being optimised for efficient measurement of radial
 velocities. The combination of resolution and signal-to-noise ratio is
 lower than is typically used for abundance determinations.  In
 addition, the spectra were obtained through small circular apertures
 (the fibers) at a wide variety of zenith distances, and with variable
 centering precison.  Thus the continuum flux distribution in the
 spectra is unreliable over wavelength ranges greater than a few tens
 of angstroms.  This combination provides a challenge for the
 determination of chemical abundances.  Jones (1991; JGW)
 developed an analysis technique
 appropriate to the derivation of true iron abundances from these
 spectra.  Full details are given in these references, with only a
 brief summary here.

 The method utilizes narrow band `photometric' indices, analogous to
 equivalent width measurements, formed by integrating the flux in a
 narrow spectral region relative to adjacent continuum regions. The
 important features of the method are that the spectral regions are
 chosen to match the resolution of the data, to contain only Fe~I lines
 which lie on the linear part of the curve of growth, and which have
 nearby regions of pseudo-continuum. The indices were identified and
 calibrated from synthetic spectra, utilising a grid of 100 synthetic
 spectra which were derived from a scaled solar model atmosphere.

 Four abundance indices were defined which measure absorption from very
 strong Fe~I lines; the pressure-sensitivity of the wings of these
 lines means that the indices are also dependent on surface gravity to
 an appreciable extent. Seven abundance indices were defined which
 measure absorption caused by weaker Fe~I lines; in the stars of
 interest to this work, these exhibit little gravity-sensitivity. Five
 indices have been identified which measure the relative strengths of
 absorption lines of ionised and neutral species. These are strongly
 sensitive to gravity.  From these 16 indices a compound iron abundance
 indicator was defined using all metallicity-sensitive indices, and the
 iron abundances derived using this compound indicator.

 Given a photometrically-determined effective temperature, then if the
 spectrum is not too noisy, S/N $\simgt 25$ in 1\AA\ pixels, it is
 possible with this technique to solve for {\it both\/} iron abundance
 and surface gravity.  This is indeed possible for the standard stars
 observed, checking the calibration of the system, but for most of the
 program stars we are forced to adopt {\it a priori\/} a value for the
 surface gravity; in this case, we assume the star is near the main
 sequence, and adopt $\log g= 4.0$. A value of $\xi=1.5$km/s for
 microturbulence was adopted for all stars.

 Stellar effective temperatures are derived from photographic V$-$I
 photometry.  A new determination of the (V$-$I) - T$_{eff}$ relation
 was made from published data, as explained in JGW. No appreciable
 metallicity dependence was found, with the adopted calibration being

 $$ { 5040 \over {\rm T}_{eff}} = 0.484 (\pm 0.010) + 0.581 (\pm 0.014)
 {\rm (V-I)}. \eqno(1)$$

 Thus for the F/G stars studied here, the typical uncertainty in
 effective temperature is $\sim 200$K.  The uncertainty in iron
 abundance resulting from given uncertainty in effective temperature
 was derived from detailed Monte Carlo simulations, and is given for
 reference in Table 1.  The uncertainties in log metallicity scale
 approximately linearly with temperature uncertainty.  Note that in
 general the estimate of the
 metallicity is more sensitive to temperature for lower
 metallicity stars, for fixed uncertainty in effective temperature, and
 at fixed signal-to-noise, for hotter stars.  A typical star in our
 sample has $T_{eff} = 5500$K and for this temperature an error of 100K
 yields an uncertainty in [Fe/H] of 0.09 dex; an error of 1.0 in $\log
 g$ yields an uncertainty of 0.15 dex in [Fe/H], and an error of 0.5
 km/s in the microturbulence parameter yields an uncertainty of 0.12
 dex at [Fe/H]$ = 0.0$, and of 0.03 dex at [Fe/H]$ = -1.5$.  Thus the
 typical uncertainty in the derived [Fe/H] may be expected to be $\sim
 0.2$dex.

 The zero-point of the iron abundance calibration was tested first by
 application to many very high S/N ratio spectra of
 the twilight sky (scattered solar light) spectrum; the
 compound indicator provided a median value of ${\rm [Fe/H] = -0.14}$, which
 was adopted as a zero-point correction.  Comparison of the iron
 abundances we derived from our observations of standard stars with
 published [Fe/H] data (mostly from Laird, Carney and Latham 1988;
 hereafter LCL) provided a mean difference, defined as the new derived
 value of this paper minus the LCL published value, of $-0.04$ dex,
 sigma $ = 0.13$ dex, for 16 stars in the range ${\rm -1.2 < [Fe/H] <
 0}$ (one extreme outlier removed).  The rms difference increased to
 0.24 dex, with a mean offset of +0.06 dex, if more metal-poor stars
 were included.  Thus the uncertainty in derived [Fe/H] expected from
 the Monte Carlo simulations is consistent with a direct comparison
 with observations.

\bigskip
 \centerline  {\it 3.2 Distances}

 Distances to the stars can be derived by photometric parallax,
 assuming the stars are in a known evolutionary stage and single
 (binarity is discussed below), given a calibration of luminosity on
 iron abundance.  We will assume the stars to be unevolved (but see
 below), and adopt the metallicity--luminosity calibration of LCL,
 which is given in terms of the offset in absolute V magnitude from the
 Hyades main sequence, for which they obtain
 $$ {\rm M_V(Hyades) = 5.64(B-V) + 0.11}.\eqno(2)$$
 The metallicity-dependent offset from
 this fiducial sequence that LCL derive (${\rm \Delta M_V^H}$) is, for a
 star of given (B--V) color and given UV excess $\delta$:
 $$ {\rm \Delta M_V^H = [ {2.31-1.04(B-V)
 \over 1.594}] [-0.6888\delta + 53.14\delta^2 -97.004 \delta^3].}\eqno(3)$$
  LCL state this calibration to be   valid for $\delta \le 0.25$,
 which equals [Fe/H]$ = -1.75$dex with the  Carney (1979)
 transformation of $\delta$ into [Fe/H], as used by LCL:
 $${\rm [Fe/H] = 0.11 -2.90\delta - 18.68\delta^2.}\eqno(4)$$
 [Fe/H]$ = -1.75$
  is the {\it mean\/} metallicity of the LCL calibrating subdwarfs; note
 that their calibrators for this relationship extend down to [Fe/H]$ =
 -2.45$ dex.  The majority of our program stars have derived
 metallicities which are well within their suggested range for this
 relationship to be valid, $> -1.75$ dex, and certainly within that of
 the calibrating subdwarfs (see below).  LCL provide an alternative
 expression for stars more metal-poor than $-1.75$dex, which they
 obtained from model isochrones.  This latter technique provides
 distances for our metal-poor stars which are in general within 10\%
 (larger) of those derived from the former, observationally anchored
 technique.  We have chosen to apply the former technique to our entire
 sample.  The typical error in our derived [Fe/H] of $\sim 0.2$dex
 leads to a random uncertainty of $\sim 20\%$ in distance estimate.
 There are also sources of systematic error.

 The possibility that the `stars' are instead  unresolved binaries means
 that the distances derived as above are underestimates, \ie\ the uncertainty
in
 distance has a systematic, as well as a random, component.  A typical
 binary system in the present sample may be taken to have a metallicity
 [Fe/H]$ \sim -0.5$ dex (see below) which with our distance estimator
 leads to a photometrically-derived absolute magnitude of the binary,
 M$_{V,P}$, given in terms of the composite color of the binary as:
 $$ {\rm M}_{V,P} = 5.31({\rm B-V})_{composite}+ 1.845. \eqno(5)$$
 Following
 Kroupa, Tout and Gilmore (1991) we can use this to express the factor
 by which the photometrically-derived distance, $d_P$, underestimates
 the true distance, $d$, in terms of the absolute V magnitudes of the
 putative binary components, M$_{V1}$ and M$_{V2}$ respectively, as :

 $$ {d_P \over d} = { (10^{-0.475 {\rm M}_{V1}} + 10^{-0.475 {\rm
 M}_{V2}})^{2.655} \over (10^{-0.4 {\rm M}_{V1}} + 10^{-0.4 {\rm
 M}_{V2}})^{3.155} }. \eqno(6)$$
 Equal-mass binaries provide the worst case,
 with the estimated distance being a factor of 1.414 ($= \surd 2$)
 smaller than the true distance. The results of Kroupa, Tout and
 Gilmore (1991; 1993) favor a high binary fraction, $\simgt 50$\%, but
 with components chosen independently from the mass function, so that
 equal masses rarely occur.  A mass
 ratio very different from unity of course means that the systematic
 errors in the distance estimator are greatly reduced from the worst
 case 40\%, and the typical errors are more likely to be much smaller.
 Indeed, for stars of the V$-$I color of the present sample, Monte-Carlo
 simulations suggest a systematic mean error of only $\sim 5$\%
   in distance  (Kroupa, Tout and Gilmore
 1993, their Figure 8).

 A further assumption that we have made which, if inappropriate, would
 systematically underestimate the true distance, is that the stars are
 all on the main sequence.  The predictions of  our star
 count models in the
 lines-of-sight of our samples are that typically $\simlt 25$\% of
 stars in the color and magnitude ranges here are evolved, with most of
 these stars being subgiants rather than red giants or red
 horizontal-branch stars.
   The error in distance estimate which results
 from assigning these stars to the main sequence obviously depends on
 color. However, the phase of subgiant evolution that has by far the
 longest duration is the initial vertical evolution (Iben 1967), when
 the star is closest to the zero-age main sequence.  The error for a
 star in this evolutionary stage is about a 20\% underestimate in
 distance.

 This systematic error applies of course only to those stars with
 colors equal to
those of the main-sequence turnoff, which is near B$-$V=0.4 for
 metal-poor stars, and B$-$V=0.55 for typical thick disk and very old
 disk stars with the abundances we derive below. In our South Galactic
 Pole data set, which is that with the largest sample, some 30\% of
 stars have B$-$V$\leq 0.60$. Fortunately, as discussed below, there is
 no significant correlation of abundance and color for stars redder
 than B$-$V=0.5, so this distance uncertainty will not affect our
 conclusions below.

 Since we do not know {\it a priori\/} which of these stars is a
 binary, or an evolved star, we should assign the above systematic
 errors to the entire sample.  Thus we expect that the distances
 derived here have about 20\% random errors, and may be
 underestimated, by perhaps 20--30\%, especially for the bluer stars in
 the sample.

 \centerline  {\it 3.3  The South Galactic Pole Sample}

 The method outlined above was applied to derive [Fe/H] abundances for
 a sample of 133 stars in the South Galactic Pole (SGP) field.  It
 should be noted that these are true {\it iron\/} abundances.  The
 observed and derived data for these stars are given in Table 2. There
 were 3 stars which were observed with enough S/N on two occasions to
 allow separate metallicity estimates, providing a (weak) internal
 check on the technique. These stars show a mean offset of $0.13$dex,
 and a dispersion of 0.2dex, in agreement with the expected
 uncertainties.  The value of [Fe/H] quoted in the table for these
 three stars is that obtained from the co-added data.

 The basic photometric data for the program stars are shown in Figure
 1(a) and (b), as V, B$-$V and V, V$-$I color-magnitude diagrams.  The
 B$-$V, V$-$I two color diagram for these stars is shown in Figure 2,
 together with the mean relation found by Reid and Gilmore (1982) from
 main sequence E-region standard stars (apparently bright and hence
 nearby and expected to have close to solar metallicity).  The large
 scatter of the data points in this last plot is due to a combination
 of photometric errors and true metallicity spread.  B$-$V is
 metallicity-dependent, being bluer for lower metallicity (for constant
 stellar T$_{eff}$), while V$-$I is primarily a measurement of
 temperature.  The metallicity dependence of the two-color diagram for
 our sample may be quantified by using the stellar evolution models of
 VandenBerg and Bell (1985; Y=0.20), and of VandenBerg (1985; Y=0.25).
 Figure 3 shows
 the colors of a one solar mass star of fixed effective temperature,
 T=5786K (chosen to minimise the interpolation required),
but a range of metallicities (and hence age, but always many Gyr).
 As can be seen, the
 models get bluer by almost 0.$^m$1 in B$-$V, with negligible change in
 V$-$I, as the metallicity is decreased from solar to one-tenth solar.

 The importance of photometric errors may be seen more clearly with a
 larger sample; Figure 4 shows the V, B$-$V data for all stars in our
 magnitude range in several $40'$ fibre fields.  The very blue stars,
 B$-$V$\simlt$0.3, occur predominantly for V $\simgt 17.3$.  As
 discussed in section 2 above, these faint magnitudes are where
 the photometric errors are becoming too large for present
 purposes (see Table 1).
   The metallicity analysis is thus restricted to only stars with
 V$< 17.30$, which incidentally also has the effect of removing all
 stars bluer than B$-$V$=0.4$, and in total removes 32 stars from
 further discussion here.  That the errors for the fainter stars are
 larger may also be seen by a plot of the B$-$V, V$-$I two color
 diagram for all 133 stars, binned by their derived metallicities, as
 shown in Figure 5(a).  The corresponding plot for only those stars
 (101 in total) brighter than V$ = 17.30$ is shown in Figure 5(b); this
 latter plot shows reduced scatter, and a trend of the offset from the
 E-region standard star line with metallicity.

 An investigation of the integrated chemical evolution of the disk
 requires a sample of stars of long enough main-sequence lifetimes to
 still be around today even if formed at early times, say 12 Gyr ago
 (\cf\ the ages derived by Edvardsson \etal 1993).  The [Fe/H] data for
 the 101 brighter stars are shown as a scatter plot against B$-$V color
 in Figure 6, together with the 12 Gyr turn-off positions from
 VandenBerg and Bell (1985; crosses) and VandenBerg (1985; asterisks).
 These isochrones provide an age of 14Gyr for the globular cluster
 47Tuc, and it is clear from the Figure that the turnoff of the bulk of
 our sample is somewhat redder than the 12Gyr isochrone, consistent
 with the inference (Wyse and Gilmore 1988; Gilmore, Wyse and Kuijken
 1989; Edvardsson \etal\ 1993) of an age of the thick disk
 comparable  to
 that of the metal-rich globular clusters.\footnote{$^2$}{Note the
 lack of metal-rich stars blueward of the theoretical turnoff
 positions. This does not imply a lack of younger metal-rich stars
 in the thin disk,
 but rather results from the magnitude-color selection. The
 stars with main-sequence lifetimes less than 12 Gyr are simply
 so luminous that to be seen with the apparent magnitude of the
 sample they would be beyond the effective edge of the thin disk
 (but not of the thick disk).
 Since  essentially {\it all\/} thin-disk stars older than 3Gyr
 have the same vertical velocity dispersion and scaleheight
 this entire range of ages will  be represented in the tail of
 the thin disk
 that is in the sample, at redder colors.}
 There remains one star that
 lies (just) blueward of the 12Gyr turnoff positions; since we are interested
 in isolating a sample of stars that have main-sequence lifetimes
 greater than the present age of the disk, we have removed this star
 from the metallicity distribution discussed below.  We will also
 restrict the sample to stars with B$-$V$ < 0.9$, further to isolate
 F/G dwarfs, which removes a further two stars.  This leaves a sample
 of 98 F/G dwarfs in our SGP sample, with 0.4$<$B$-$V$<$0.9.

 Even just considering the stars brighter than V$ = 17.30$, comparison
 of Figures 1 and 4 shows that there is a clear deficiency of stars
 bluer than B$-$V$ = 0.5$ in the sample for which iron abundances were
 derived, compared to the available stars on the sky.  This arises
 since it is more difficult to derive reliable abundance estimates for
 these hotter stars from the available low S/N data.  This bias is
 important since it is clear from the data that the metal-poor stars
 tend to be bluer, and thus the effective temperature bias will
 translate into a metallicity bias.  This may be corrected for by
 renormalising the color distribution of stars in the metallicity
 sample to agree with that available on the sky.  The number in the
 metallicity sample relative to the area-complete sample, in color bins
 of 0.1 mag starting at B$-$V$ < 0.5$ and ending at B$-$V $=0.9$ is
 0.35; 0.68; 0.58; 0.60 and 0.73 respectively.  This means that,
 allowing for the limitations of small number statistics, these ratios
 are 0.5:1:1:1:1 and the metallicity distribution needs to be corrected
 only for those stars in the bluest B$-$V bin, and that the stars in
 that bin should be given double weight in the iron abundance
 distribution.  The resulting distribution is then free of color bias.
 The metallicity of the initial, biased sample is shown as the dashed
 histogram in Figure 7, together with the color-corrected distribution
 (solid histogram).  The correction is clearly a small effect.

 The iron-abundance distribution also needs to be corrected to derive a
 volume-complete sample from our area-complete photometrically-defined
 sample.  Distances to the stars were derived as described above.  The
 scatter plot of [Fe/H] against distance for the 98 stars with V$ <
 17.30$ and 0.4$<$B$-$V$<$0.9 is shown in Figure 8.  The scatter plot
 of B$-$V versus distance for these stars is shown in Figure 9,
 together with the limits set by our pruned magnitude range, for given
 metallicity, with the solid lines corresponding to the faint limit of
 the sample and the dashed lines to the bright limit.  The sample is
 complete for a given iron abundance within the volume defined by the
 two distance limits.  It can be seen from combination of Figures 8 and
 9 that the more metal-rich stars observed tend to be found
 preferentially near the inner distance limit of the sample selection
 criteria.  This is despite the fact that were metal-rich stars to
 exist at large distances, they would be favored by the selection in
 magnitude and color.  Thus one may infer immediately that the more
 metal-poor stars are in a more spatially-extended distribution than
 the more metal-rich stars.

 The metallicity distribution of a volume-complete sample may be
 derived from the observed distribution (after the correction for color
 bias in the metallicity determinations described above) by calculating
 the volume accessible to a given metallicity bin at its characteristic
 distance, and then bringing all metallicity bins to a common, fiducial
 distance from the plane.  Note that we prefer the method of direct
 integration to that of V/V$^{'}_{max}$; the V/V$^{'}_{max}$ method is less
 appropriate here as there is a mix of stellar populations with
 overlapping abundance distributions but different spatial
 distributions.  In our simplest case below all stars are assumed to be
 thick-disk stars, and our direct integration method and the
 V/V$^{'}_{max}$ method are equivalent.

 The volume accessible to each metallicity bin is the integral of the
 volume element between the maximum and minimum distances shown in
 Figure 9, achieved by first integrating at a fixed color, and then
 simply summing over color, since corrections for color bias have
 already been applied. The characteristic distance for each metallicity
 bin is obtained by integrating the line element, weighted by the
 density profile of the stars, between the limits set by the color
 selection; this is necessary since the stars do not uniformly occupy
 the accessible volume.  Thus some model-dependency enters through
 choice of density profile; use of the individual distances derived
 above to determine a mean distance is hampered by the small numbers in
 the metal-poor and metal-rich ranges of the sample. 	 The present
 small sample is obviously not well-suited for the determination of
 self-consistent density profiles.  We adopted the thin disk and thick
 disk scaleheights obtained by Kuijken and Gilmore (1989) for K dwarfs
 at distances comparable to those probed by the present sample,
 300-4000pc, {\it viz.} 250pc for the thin disk and 1000pc for the
 thick disk, and have modelled the stellar halo by an exponential
 profile with scaleheight 4000pc, much larger than either of the disks.

 As discussed in section 2 above,
 the present sample is expected to be dominated by thick disk
 stars, with some contribution from the stellar halo presumably mostly
 at the metal-weak end, and from the thin disk at the metal-rich end.
 As noted above, the distance distribution of the sample does imply a
 more extended spatial distribution for the more metal-poor stars.  We
 tested the sensitivity to the density profiles by calculating the
 corrections in two ways -- first by simply adopting the thick disk
 density profile for all metallicity bins, and secondly by adopting the
 stellar halo density profile for stars more metal-poor than $-1$dex,
 the thick disk density profile for all stars above this until
 $-0.4$dex, allowing 10\% thin disk contribution to the stars with
 $-0.4 < $[Fe/H]$ < 0$, and assigning all stars more metal rich than
 solar to the thin disk. These percentages are
in broad agreement with the expectations from star-count models.
In general one expects the results to be
 rather insensitive to the choice of density profile except for the
 most metal-rich bins, where the thin disk density profile is applied
 and the depth of the sample is a significant number of scaleheights.
 This expectation is borne out by Table 3, which gives the resultant
 relative numbers of stars per unit volume at a fiducial distance of
 1500pc, chosen to be close to the middle of the range of distances
 sampled by the data.  The histograms of the volume-complete
 metallicity distributions obtained these two ways are shown in Figure
 10.  Note that the metallicity distribution of the thin disk is of
 course better constrained by a sample more concentrated to the plane;
 such is the local Gliese catalog sample analysed in the companion
 paper.  Indeed, both the metal-poor bins, [Fe/H]$< -1$, and the
 metal-rich bins, [Fe/H]$>$ solar, are based on rather a small number
 of stars, so even although they are apparently well-populated in the
 corrected distributions, their error bars are large.

\bigskip
 \centerline  {\it 3.4 The Rotation Field}

 Iron abundances were obtained for a sample of 87 stars in the
 direction of $(\ell,b)=(270,-45)$, UK Schmidt Field 117, as tabulated
 in Table 4.  There were 3 stars which were observed with enough S/N on
 two occasions to allow a metallicity estimate to be made, providing an
 internal check on the calibration.  This yielded a mean offset of 0.05
 dex, again with a dispersion of 0.2dex, and again the value quoted in
 the table is that obtained from the co-added data.

 Following the analysis of the SGP stars, the V, B$-$V color magnitude
 diagram of the sample with derived iron abundances is shown in Figure
 11(a), and the V, V$-$I data are shown in Figure 11(b).  Note the few
 stars brighter than V=16 which were observed as a poor-seeing back-up;
 these were drawn at random from the same color range as the fainter
 stars.  The V, B$-$V data for representative entire fibre fields
 are shown in Figure 12.  Again, there is a
 deficiency of blue stars, B$-$V $\simlt 0.55$, in the metallicity
 sample. Although there is no trend of [Fe/H] with B$-$V color, for
 B$-$V $\simgt 0.6$ where we have a large enough sample in this field,
 comparison with the SGP data suggests that there may be an effect for
 bluer stars, and lower metallicities [Fe/H]$\simlt -1$dex, and we
 would not be aware of it here.  The metallicity distribution may be
 unaffected by the color bias, but conservatively one should treat the
 derived number of stars in this field with [Fe/H]$\simlt -1$dex as
 lower limits only (as we discuss below, the fact that there is
 little difference between the distribution in the SGP and that
 found in this field suggests that any remaining bias against
 metal-poor stars is a small effect).
 For convenience, we note that binning the data
 into $0.^m1$ ranges of B$-$V, centered on 0.55, 0.65, 0.75 and 0.85,
 the ratios of the number of stars in the metallicity sample to that in
 the entire fibre field are 0.12; 0.20; 0.20; 0.10.  Since there is no
 trend between color and metallicity in this range the lack of
 constancy of this ratio is not meaningful in determining the derived
 metallicity distribution in this field.  We also for consistency
 remove stars with B$-$V$\ge0.9$, which leaves a total of 83 stars in
 this line-of-sight.

 The two color diagram for the stars in this field with metallicity
 determinations is shown in Figure 13.  The corresponding diagrams for
 sub-samples selected by iron abundance is shown in Figure 14, where
 there is a clear trend in offset from the E-region standard star ridge
 line with decreasing metallicity.  The photometry in all fields is
 calibrated independently, as discussed in section 2 above.  Unlike the
 in SGP, there are no indications in this field that the errors in the
 photometry are unacceptable for the faint stars.
   However, it should be noted that the
 most metal-poor stars are faint, and hence the derived metallicities
 have larger error bars. In compensation, these bins are sparsely
 populated and are given low weight in the discussion below.

 Following the analysis for the SGP stars, Figure 15 shows the scatter
 plot of B$-$V color versus [Fe/H], together with the 12 Gyr turnoff
 positions from VandenBerg and Bell (1985) and from VandenBerg (1985).
 There are no stars in this field which should be excluded by virtue of
 their location in this plot.  The color and magnitude limits for this
 field are then $15.0 \le {\rm V} \le 17.87$, $0.5 \le {\rm B-V} <
 0.9$; as discussed in section 2 the coverage  of stars within these limits
 is not uniform.

 The distances for the F117 stars were derived as for the SGP stars.
 The direction of this field is such that the planar radial
 Galactocentric distances of the F117 stars are not significantly
 different from the solar circle -- the maximum planar Galactocentric
 distance probed by the sample is 8.6kpc, for a solar Galactocentric
 distance of 8kpc.  Most stars are at distances from the Sun of 1.5kpc,
 corresponding to a planar Galactocentric distance of 8.1kpc.
 Comparison of the iron-abundance distribution in this field with that
 of the SGP is therefore not sensitive to the presence or otherwise of
 radial Galactic abundance gradients.  The vertical height distribution
 of the sample in this field covers the range $0.7 \simlt z {\rm (kpc)}
 \simlt 1.5$, with approximately equal numbers below and above $z
 =1.1$kpc (since $z = d \times \sin 45^o$ the vertical distances are
 trivially derived from the distances, $d$, given in the Table).  The
 metallicity distribution in this field, when combined with the SGP
 data above which have equal numbers of stars above and below 1650pc,
 thus provides a probe of the vertical metallicity gradient of the
 thick disk at the solar circle.

 Figure 16 shows the stars in the [Fe/H], distance plane, and Figure 17
 the B-V, distance plane, where the lines indicate the distance limits
 corresponding to the sample magnitude limits for given metallicity.
 Corrections for the variation of the volume sampled with color and
 metallicity were made in the same way as the analysis of the SGP
 sample. Table 5 gives the binned iron-abundance distribution for the
 `raw' sample, together with characteristic distances and number of
 stars per unit volume, at a fiducial vertical distance of 1000pc,
 under the two sets of assumed density profiles.  Histograms of these
 iron-abundance distributions are shown in Figure 18.  Again, the most
 robust determination is for the iron-abundance distribution in the
 range $-1.0 < {\rm [Fe/H]} < 0$, with the (noisy) data below [Fe/H]$=
 -1$ being formally a lower limit to the numbers of these stars (although as
discussed in section 4 below, this is not in fact a problem
 in practice).

 \bigskip
 \centerline {4. DISCUSSION}
 \bigskip

 The abundance distribution at a distance of 1.0-1.5kpc from the
 Galactic Plane which we have derived above contains valuable
 information on the star-formation and chemical-enrichment
 history of the thick disk. In a
 complementary  paper (Wyse and Gilmore 1994) we combine this distribution
 with a new determination of the abundance distribution in the Galactic
 Plane to reconsider the evolution of the Galactic disk at the solar
 Galactocentric radius. In a future paper we combine the present data
 with our larger kinematic and photometric survey and other relevant
 published data.  Prior to those analyses, we note here some of the
 more obvious implications of our thick disk abundance distribution.

\bigskip
 \centerline  {\it 4.1 Is the Thick Disk Old?}

 The combination of chemical abundance and photometric data allows
 determination of turnoff ages.
 The data and isochrones shown in
 Figure 6 for the SGP and Figure 15 for F117 are both confirmatory
 evidence that the thick disk is predominately composed of stars which
 are as old as the oldest disk stars, and possibly as old as the metal-rich
 globular clusters (\cf\ Nissen and Schuster 1991).

 Most models of thick-disk formation suggest that there should be an
 age spread in this component. This will reflect the star-formation
 history of the precursor thin disk and/or satellite for slow or rapid
 kinematic heating models. Star formation in a pressure-supported
 thick gas disk, or models in which there is a burst of star formation during
 a merger event can however produce a narrow age range (although
 merger-induced star formation may be expected to be rather more
 centrally-concentrated than the thick disk is observed to be, due
 to gravitational torque-induced gas flows
 towards the central regions; Mihos and Hernquist 1994).
 Our present data are insufficient to investigate such possibilities,
 as isochrone crowding implies that high-precision, narrow-band photometric
 data are required for adequate age resolution at old ages. The
 predominant old age shown in these data does  however limit all formation
 models to have the creation of the thick disk at early times. Thus, models
invoking slow
 kinematic heating of thin-disk stars must appeal to special heating
 conditions in the early Galaxy, while merger models must have only
 early merger(s), or produce thick disks that consist predominantly
 of stars from the shredded satellite, which must itself be
 predominantly old.

 More detailed element-ratio data allow
 determination of star-formation histories, initial mass functions, and
 interstellar-medium mixing timescales and efficiencies.  These topics
 will be considered in some detail in a later paper.  However, the fact
 that thick-disk stars have element ratios that reflect incorporation
 of iron from Type I supernovae (\eg\ data of Edvardsson
 \etal\ 1993) implies a formation timescale more extended than the timescale
 appropriate for
 these supernovae, probably
 $\sim$1Gyr.  Indeed, the Edvardsson \etal\ estimates of stellar
 ages directly suggest a range of ages for the thick disk at the
 solar neighborhood (defined by the mean radius of their orbit
 calculated in Edvardsson \etal\ as being between 7kpc and 9kpc)
 of at least a few Gyr, $\simgt 12$Gyr (note that the kinematic
 selection effects in that sample make it difficult to isolate
 thick-disk stars at other radii).
 It may also be noted that the Edvardsson \etal\ age estimates  do not show
 any significant age gap between the thick disk and thin disk
 stars, as would be expected if one were to believe the
 combination of the (young) white dwarf ages for the thin disk and
 the (old)  stellar
 evolutionary ages for the globular clusters.  Further, their
 element ratio data do not show any indication of a hiatus in star
 formation -- if there were a hiatus,   then
 enrichment of oxygen, and of other alpha-elements which are primarily
 produced in Type II supernova exposions of massive stars, would
 cease for the duration of the hiatus,
 but enrichment in iron would continue, in line with current models of  Type I
 supernovae  (binary white dwarfs) which allow for an final
 explosion many Gyr after the birth of the progenitors.
 Stars born after a hiatus in star formation are then formed from
 gas which has been
 preferentially enriched in iron, and in particular will show low
 oxygen-to-iron ratios (Gilmore and Wyse 1991).  However, the
 element ratio data of Edvardsson \etal\ are continuous at the
 solar neighborhood, as can be
 seen from their Figure 21 (alpha-to-iron versus age) and their
 Figure 22 (oxygen-to-iron ratio versus iron abundance).

\bigskip
 \centerline  {\it 4.2 Is There a Vertical Abundance Gradient in the Thick
Disk?}

 A comparison between the iron-abundance distribution in the SGP (at a
 fiducial distance of 1500pc) and in F117 (at $ z=1000$pc) is shown in
 Figure 19.  It is clear that there is no significant difference
 between the distributions in these two fields. Further, this
 comparison suggests that the incompleteness for  [Fe/H]$ < -1 $dex in
 F117 cannot be a large effect.
   The thick disk abundance distribution in the Galactic
 Plane is poorly determined, but in so far as it is known may be
 characterised as a Gaussian with mean abundance at $-0.6$dex (Carney,
 Latham and Laird, 1989), again consistent with the {\it in situ} data
 presented here. Thus, the available thick disk abundance
 determinations as a function of distance may most easily be
 interpreted in terms of a constant iron-abundance distribution for the
 thick disk, independent of height above the plane, at least over the
 range probed by these data.  This is in agreement with the conclusions
 from  available {\it in situ\/} samples of thick-disk stars with
 UV-excess photometric metallicity estimates (Gilmore and Wyse 1985; Majewski
 1992).

 In principle the absence of a significant abundance gradient is weak
 evidence against some models of thick disk formation. The relevant
 models may crudely be categorised as slow dissipative formation from
 gas with the same spatial distribution as the resulting thick stellar
 disk; slow kinematic heating of stars formed in a thin disk, or
 disruptive or very rapid heating of a thin disk.  Scenarios of thick
 disk formation that invoke disruptive heating of a pre-existing thin
 disk, by means such as satellite galaxy accretion, will in general
 predict lower amplitude metallicity gradients in the thick disk than
 do scenarios that invoke slow dissipative settling.  However, if there
 were a vertical abundance gradient in a thin disk before rapid
 heating, then current models suggest that dissipationless relaxation
 after merging will weaken such a metallicity gradient, but not erase
 it (White 1980; Hernquist and Quinn 1993), so there is not a clear
 dichotomy between predictions of these two scenarios.

 Were the thick disk a tracer of a fairly major merging event --
 accretion of a satellite of mass around 10\% or more of the mass of
 the Galactic disk (Quinn, Hernquist and Fullager 1993) -- then one may
 expect two places of origin of the stars that are now in the thick
 disk, some being shredded satellite, others being originally thin
 Galactic disk.  Satellite galaxies presently observed appear to follow
 a well-defined luminosity-metallicity relationship that is not a
 simple extrapolation of that of normal galaxies (Caldwell \etal\
 1992).  It may not be coincidence that the mean metallicity of the
 thick disk, $\sim -0.6$dex, is comparable to that of the Small
 Magellanic Cloud today, which is probably of sufficient mass to be
 representative of the type of satellite that could cause a thick disk
 to form when accreted.  Satellite galaxies of the type which surround
 the Milky Way today typically have metallicity of $\sim -1.5$dex and
 absolute magnitudes $M_V \sim -12$ (Armandroff \etal\ 1993). Such a
 galaxy would probably not have sufficient impact to create a thick
 disk by heating, but could contribute stars by direct shredding. Of
 course, what is really relevant is the stellar mass of the thin disk
 perhaps 12Gyr ago, and the chemical abundance of the stars in the
 satellite galaxy at that epoch. Merger models clearly are able to
 generate complex age and abundance distributions in the thick disk. A
 common feature is however a weak or zero vertical abundance gradient,
 consistent with present observations.

 If the thick disk were formed from the thin disk in a manner which
 preserves the initial conditions in the thin disk then one might
 expect to observe a vertical abundance gradient.  For example, a model
 which invokes slow kinematic heating of a thin disk to create the
 thick disk will produce a thick disk which retains memory of any
 age-metallicity relation which may have been established in the thin
 disk. Given that there is little or no evidence for any
 age-metallicity relation in the thin disk today, it remains unclear if
 this property of slow heating models provides an observational
 signature in the abundance gradient. Such models are better tested
 from age data.

\bigskip
 \centerline  {\it 4.3 The Abundance Distribution Function at z=1.5kpc}

With the knowledge that the F117 metallicity distribution is indeed
complete at the metal-poor end, and that there is no evidence for a
significant vertical abundance gradient in the data, we are now able
to combine the SGP and F117 abundance distributions into our final
{\it in situ} abundance distribution.  We first correct the F117
distribution function to z=1.5kpc, for consistency with the SGP data.
We do this for both the `every star is a thick-disk star' density
profile model and for the three-populations model. The distributions
in the two fields do not differ significantly in either shape or
number of stars in the original sample, so we can combine them by
summing. The resulting distribution functions are presented in Table
6, which represents our final abundance distribution function. This
distribution function is combined with an updated thin-disk
metallicity distribution derived from the Gliese catalog, and the
composite analysed, in a separate paper. We now consider the more
metal-poor stars.

The present sample has been highly optimised for study of the {\it in
situ\/} thick disk. Consequently, these data are not ideal for a study
of either the thin disk or the stellar halo.  However, one can use
information on these latter populations from other sources to aid in
the interpretation of the iron abundance data of the present sample.
In particular, the contribution of the stellar halo to the metal-poor
stars can in principle be estimated by combination of its chemical
abundance distribution and the predictions of star-count models,
allowing statistical isolation of the metal-poor stars in the thick
disk from those in the halo.

Extension of the metallicity distribution function of the thick disk
to low metallicities is a natural expectation in most scenarios for
thick-disk formation.  For example, should the thick disk represent
the first stars to form in the disk, then one expects stars of
arbitrarily low metallicities in the thick disk.  The thick disk and
thin disk are in fact plausibly chemically-connected from the
continuity of element ratios and ages in the data of Edvardsson \etal\
(1993).  The kinematics of thick and thin disk also suggest an
evolutionary connection -- as seen in the similarity of their
angular-momentum distributions (Wyse and Gilmore 1992) and in the
locations of thick- and thin-disk stars in the the orbital
eccentricity--metallicity plane (most recently in Twarog and
Anthony-Twarog 1994). Thus it is plausible that the thick disk is the
chemical precursor to the thin disk, and that the earliest, and most
metal-poor, disk stars to form in the Galaxy are now in the thick
disk.

The number of such very metal-poor stars, or more correctly the shape
and normalisation of the metal-weak tail to the thick disk (and the
thin disk, for that matter) cannot be predicted reliably from first
principles. Indeed, this difficulty is simply a restatement of the
`G-dwarf problem'.  Depending on details of such things as
pre-enrichment from previous stellar generations, gas flows,
inhomogeneities {\it etc.\/}, one may have any distribution from that
of the Simple Model, which has a substantial metal-weak tail, to a
distribution function with a severe G-dwarf problem, which has very
few metal-poor stars.  Further, as mentioned above, merger/satellite
disruption models are likely to generate a metal-weak tail to the
thick disk, containing stars shredded from the satellite and accreted
during the merger, in addition to any metal-poor stars in the original
disk.

Note however that substantial pre-enrichment, which is the easiest way
to reduce the expected number of metal-poor stars, appears unlikely.
The low angular momentum of the stellar halo points to the central
bulge as the recipient of gas ejected during the evolution
of this component (Eggen, Lynden-Bell and Sandage 1962; Carney, Latham
and Laird 1990; Wyse and Gilmore 1992) rather than the disk, and thus
any such metal-enriched gas is not available to `pre-enrich' the thick
disk.  Predictions of the expected number of metal-weak thick disk
stars are clearly very model-dependent, though some are expected in
almost all models.

The metallicity distribution of the stellar halo has been defined by
many samples, some kinematically selected, some not, with a concensus
that it is adequately described mathematically by a Gaussian in log
metallicity, of mean $-1.5$dex and of sigma 0.4dex (e.g.~Norris and
Ryan 1989; Laird \etal\ 1988).  The iron abundance distributions of
the present paper indeed shows a local maximum at $-1.5$dex.  The
kinematically-defined local thick disk has a metallicity distribution
that can also be mathematically described by a Gaussian, of mean
$-0.6$dex and of sigma 0.3dex (\eg\ Carney \etal\ 1989).  Several
studies have identified stars with kinematics that assign them to the
thick disk, but with [Fe/H]$ \simlt -1$dex (\eg\ Norris, Bessel and
Pickles 1985; Morrison, Flynn and Freeman 1990; Morrison 1993a).  What
fraction of the metal-poor stars seen in the present sample can one
ascribe to the thick disk and to the halo?

Even with no chemical abundance gradients, the expectation will depend
on the distance of the sample under study, due to the different
density profiles of the thick disk and of the halo.  Locally, the
relative normalisations of thick disk and halo to the thin disk are
$\sim 0.04/0.00125$ or around a factor of 30 more thick disk stars
than halo stars.  To within the accuracy of the simple Gaussian model
fits to the iron abundance distributions, [Fe/H]$ = -1$dex is
equidistant in terms of sigma from the means of the thick disk and of
the halo.  THus combination with the local normalisation gives that at
this metallicity the relative numbers of thick disk and halo stars
locally should be around 30, or that the vast majority of local stars
with [Fe/H]$ \sim -1$dex should be associated with the thick disk
rather than with the halo.  For  a
typical scaleheight of  1kpc for the thick disk and
an effective scaleheight of 4kpc for the halo,
this ratio decreases to  10 at 1.5kpc.

Scaling from simple local normalisations and scaleheights to compare
with the present number per unit volume at 1500pc is complicated due
to the significantly different color--magnitude relations of the
different stellar populations, which are most different in the
color--apparent magnitude range selected.  The direct predictions from
star-count models for the actual observed sample are, as discussed in
section 2 above, that the sample, as defined in color--apparent
magnitude space, should contain $\sim 50\%$ thick disk and $\sim 30\%$
halo, or a ratio of 5:3 thick disk:halo.

The observed (but color-bias-corrected) SGP iron abundance
distribution, as given in column 2 of Table 3, may be used to estimate
the thick disk:halo ratio using the Gaussian fits, applied only over
the range $\pm 1\sigma$.  Here the observed local maximum at $-0.6$dex
is rather broad, but taking the sum of three bins centered either on
$-0.5$dex or on $-0.7$ dex (which is $\pm 1\sigma$ for the Gaussian
fit to the local thick disk) gives $\sim 50$ stars.  Summing three
bins centered on $-1.5$dex, together with half of the next bins on
either side, to give $\pm 1 \sigma$ for the halo, results in 10 stars.
Thus assigning stars on the basis of an assumed Gaussian
characterisation of the thick disk and halo metallicity distributions
suggests a ratio of $\sim 5:1$.  This is a factor of $\sim 3$ {\it
fewer\/} halo stars than predicted by the star-count models
and a
factor of $\sim 2$ {\it more} halo stars than is calculated from the
simple Gaussian fitting argument above.

Admittedly the root-N uncertainty on this ratio is rather large, but
the discrepancy between observations of halo stars and the star-count
predictions is in broad agreement with that found by others (\eg\
Friel 1987, 1988; Morrison 1993b).  The breakdown of the star-count
predictions could be much worse than this, allowing no quantitive
statement to be made about the metal-poor stars in the thick disk.  It
is interesting however that there is increasing evidence both that the
number of halo stars seen far from the plane is not consistent with
expectations from commonly adopted halo number density normalisations
and density profiles, and that the simple Gaussian description of
abundance distribution functions is invalid.  The kinematics of the
stars will help provide a more definitive analysis, which is deferred
to a future paper.

\vfill\eject

\centerline  {ACKNOWLEDGEMENTS}
\bigskip
The Center for Particle Astrophysics is supported by the NSF.
RFGW acknowledges support from the AAS Small Research Grants Program
in the form of a grant from NASA administered by the AAS, from the NSF
(AST-8807799 and AST-9016226) and from the Seaver Foundation.  Our
collaboration was aided by grants from NATO Scientific Affairs
Division and from the NSF (INT-9113306).  GG thanks Mount Wilson and Las
Campanas Observatories for access to their excellent facilities, as a
Visiting Associate, during the early stages of this work.

\bigskip
\centerline  {REFERENCES}
\parindent=0pt
\def\apjl{ApJL}
\bigskip

\pp Aguilar, L., Carney, B., Latham, D. and Laird, J. 1994, AJ (in
press)

\apjref Armandroff, T., Da
Costa, G. S., Caldwell, N. and Seitzer, P. 1993;\aj;106;986

%\apjref Audouze, J. and Tinsley, B. 1976;\araa;14;43

%\apjref Baldwin, J.A. etal 1991;\apj;374;580

\apjref Bergbusch, P.A. and VandenBerg, D.A. 1992;\apjs;81;163
\apjref Bergh, S. van den 1962;\aj;67;486
%\pp Bergh, S. van den 1993. in IAU Colloquium 145, `Supernovae and Supernova
%Remnants' in press.

%\pp Bessell, M. S. 1979, PASP, 91, 589.

%\pp Bikmaev, I.F., Bobritskii, S.S. and Sakhibullin, N.A. 1990, Sov Astr Lett,
%16, 91.

%\pp Blitz, L. 1990, in `The Evolution of the Interstellar Medium',
%ed L.~Blitz, ASP Conf. Ser. {\bf 12} (ASP: San Francisco) p273.

\apjref Burstein, D. and Heiles, C. 1982;\aj;87;1165
\apjref Caldwell, N., Armandroff, T. E., Seitzer, P. and Da Costa, G.
S. 1992;\aj;103;840

\apjref Carney, B. W., 1979;\apj;233;211
%\apjref Carney, B. W., 1983a;\aj;88;610
%\apjref Carney, B. W., 1983b;\aj;88;623
\apjref Carney, B.W., Latham, D. and Laird, J. 1989;\aj;97;423
\apjref Carney, B.W., Latham, D. and Laird, J. 1990;\aj;99;572

\apjref Cayrel, R., Perrin, M.-N., Barbuy, B. and Buser, R.
1991a;\aap;247;108

\apjref Cayrel, R., Perrin, M.-N., Barbuy, B., Buser, R. and
     Coupry, M.-F. 1991b;\aap;247;122

%\apjref Clarke, C.J. 1991;\mnras;249;704
%\apjref Crawford, D.L. and Barnes, J.V. 1969;\aj;74;407
%\apjref Cunha, K. and Lambert, D.L. 1992;\apj;399;586
%\apjref Edmunds, M.G. 1975;Ap Sp Sci;32;483
\apjref Edmunds, M.G. 1990;\mnras;246;678
\apjref Edvardsson, B., Andersen, J., Gustafsson, B., Lambert, D.L., Nissen,
P. and Tomkin, J. 1993;\aap;275;101

\apjref Eggen, O., Lynden-Bell, D. and Sandage, A. 1962;\apj;136;748
%\pp Elmegreen, B.
%1990, in `The Evolution of the Interstellar Medium',
%ed L.~Blitz, ASP Conf. Ser. {\bf 12} (ASP: San Francisco) p247.

\apjref Friel, E.D. 1987;\aj;93;1388
\apjref Friel, E.D. 1988;\aj;95;1727

%\pp Friel, E. 1993, in `The Globular Cluster -- Galaxy Connection', eds
%G.~Smith and J.~Brodie, ASP Conf Series {\bf 48},
%(ASP: San Francisco) p273.

%\apjref Gerola, H. and Seiden, P.E. 1978;\apj;223;129

%\pp de Geus, E. 1991, in `The Formation and Evolution of Star
%Clusters', ed K.~Janes, ASP Conf Series {\bf 13}, (ASP: San Francisco) p40.

\pp Gilmore, G. 1984,  in Astronomy with Schmidt-type Telescopes,
edited by M.~Capaccioli (Reidel, Dordrecht) p77

\apjref Reid, I.N. and Gilmore, G. 1982;\mnras;201;73

\apjref Gilmore, G. and Reid, I.N. 1983;\mnras;202;1025

\apjref Gilmore, G. Reid, I.N. and Hewett, P.C. 1985;\mnras;213;257

\apjref Gilmore, G. and Wyse, R.F.G. 1985;\aj;90;2015
\apjref Gilmore, G. and Wyse, R.F.G. 1986;Nature;322;806

\apjref Gilmore, G., Wyse, R.F.G. and Kuijken, K. 1989;\araa;27;555

\pp  Gunn, J.E. 1987 in {The Galaxy}, edited by
G.~Gilmore and B.~Carswell (Reidel, Dordrecht) p413

%\apjref Gustafsson, B. and Nissen, P.E., 1972;\aap;19;261

%\apjref Hartkopf, W.I. and Yoss, K.M. 1982;\aj;87;1679
%\apjref Hartwick, F.D.A. 1975;\apj;209;418

%\apjref Herbig, G. and Tenrdrup, D.M. 1986;\apj;307;609
\pp Hernquist, L and Quinn, P.J. 1993 in Galaxy Evolution: The Milky
Way Perspective, edited by  S.~R.~Majewski (ASP, San Francisco) p187
\apjref Iben, I. 1967;\araa;5;571
\pp Jones, J.B., 1991, Ph.D.~thesis, Univ of Wales, Cardiff

\pp Jones, J.B.,  Gilmore, G. and Wyse, R.F.G. 1994 (in preparation)

%\apjref Kaufmann, M. 1975;Ap Sp Sci;33;265
%\apjref Keenan, F.P., Hibbert, A. and Dufton, P.L. 1985;\aap;147;89

\apjref Kroupa, P., Tout, C.A. and Gilmore, G. 1991;\mnras;251;293
\apjref Kroupa, P., Tout, C.A. and Gilmore, G. 1993;\mnras;262;545

\apjref Kuijken, K. and Gilmore, G. 1989;\mnras;239;605

%\pp Lada, C.J. and Lada, E.A. 1991, in `The Formation and Evolution of Star
%Clusters', ed K.~Janes, ASP Conf Series {\bf 13}, (ASP: San Francisco) p3.

\apjref Laird, J., Carney, B. and Latham, D. 1988 (LCL);\aj;95;1843
\apjref Laird, J., Rupen, M.P., Carney, B. and Latham, D. 1988;\aj;95;1908

%\apjref Lewis, J.R. and Freeman, K.C. 1989;\aj;97;139

%\apjref Lynden-Bell, D. 1975;Vistas Ast;19;299

%\apjref Magain, P., 1987a;\aap;179;176
%\apjref Magain, P., 1987a;\aap;181;323
\apjref Majewski, S.R. 1992;\apjs;78;87
%\pp Malinie, G., Hartmann, D.H. and Mathews, G.J.
%1991, in `The Formation and Evolution of Star Clusters',  ed K.~Janes
%ASP Conf
%Series {\bf 13}, (ASP: San Francisco) pXX.

%\apjref Malinie, G., Hartmann, D.H., Clayton, D.D. and Mathews, G.J.
%1993;\apj;413;633

%\apjref Matteucci, F. and Fran\c cois, P. 1989;\mnras;239;885
\apjref Mihos, J.C. and Hernquist, L. 1994;\apjl;425;L13
%\apjref Monet, D. \etal 1992;\aj;103;638

\apjref Morrison, H., 1993a;\aj;105;539
\apjref Morrison, H., 1993b;\aj;106;578
\apjref Morrison, H., Flynn C. and Freeman K.C., 1990;\aj;100;1191

\apjref Neese, C.L. and Yoss, K. 1988;\aj;95;463
%\apjref Nissen, P.E. 1970a;\aap;6;138
\apjref Nissen, P.E. and Schuster, W. 1991;\aap;251;457
%\apjref Nissen, P.E. 1970b;\aap;8;476
\apjref Norris, J. and Ryan, S., 1989;\apj;340;739
\apjref Norris, J. and Ryan, S., 1991;\apj;380;403
\apjref Norris, J., Bessel, M. and Pickles, A. 1985;\apjs;58;463
%\apjref Olive, K.A. and Schramm, D.N. 1982;\apj;257;276
%\apjref Olsen, E.H. 1983;\aaps;54;55
%\pp Pagel, B.E.J. 1989, in `Evolutionary Phenomena in Galaxies',
%eds J.E.~Beckman and B.E.J.~Pagel (CUP: Cambridge) p201.

\pp Perrin, M.-N., Friel, E., Bienayme, O., Cayrel, R., Barbuy,
B. and Boulon, J. 1994, \aap (in press)

%\apjref Pagel, B.E.J. and Patchett, B.E. 1975;\mnras;172;13
%\apjref Peterson, R. C., and Carney, B. W., 1979;\apj;231;762
\apjref Quinn, P.J., Hernquist, L. and Fullagher, D., 1993;\apj;403;74
\apjref  Rana, N.C. 1991;\araa;29;129

%\apjref Reeves, H. 1972;\aap;19;215
\apjref Reid, I.N. and Gilmore, G. 1982;\mnras;201;73
%\apjref Rubin etal 1991;\apj;374;564
%\apjref Ryan, S.G. and Norris, J.E. 1991;\aj;101;1865
\pp Ryan, S.G. and Norris, J.E.  1993, in Galaxy Evolution: The Milky
Way Perspective, edited by  S.~R.~Majewski (ASP, San Francisco) p103

%\apjref Sandage, A., 1965;\apj;00;00
%\apjref Saxner, M. and Hammarb\"{a}ck, G., 1985;\aap;151;372

\apjref Schmidt, M. 1963;\apj;137;758
%\apjref Schuster, W.J. and Nissen, P.E. 1989a;\aap;221;65
\apjref Schuster, W.J. and Nissen, P.E. 1989;\aap;222;69
%\pp Searle, L. 1972, in `L'Age des Etoiles', IAU Colloquium 17,
%eds G.~Cayrel de Strobel and A.M.~Delplace (Obs.~de Paris-Meudon:
%Paris) paper 52.

%\apjref Searle, L. and Zinn, R. 1978;\apj;225;357

%\apjref Shaver, P.A., McGee, R.X., Newton, L.M., Danks, A.C. and Pottasch,
%%S.R.
%1983;\mnras;204;53
\apjref Silk, J. and Wyse, R.F.G. 1993;Phy Rep;231;293

\apjref Stobie, R., Gilmore, G. and Reid, I.N. 1985;\aaps;60;495

\apjref Stobie, R., Sagar, R.  and Gilmore, G. 1985;\aaps;60;503

%\apjref Talbot, R.J. and Arnett, W.D. 1973;\apj;186;69

%\apjref Talbot, R.J.  1973;\apj;186;69
%\apjref Tinsley, B. 1975;\apj;197;159
\apjref Tinsley, B. 1980;Fund Cosmic Phys;5;287
\apjref Twarog, B.A. and Anthony-Twarog, B.J. 1994;\aj;107;1371
\apjref VandenBerg, D.A. 1985;\apjs;58;711
\apjref VandenBerg, D.A. and Bell, R.A. 1985;\apjs;58;561

\apjref White, S.D.M. 1980;\mnras;191;1P
%\apjref Warren, W.H. and Hesser, J. 1978;\apjs;36;497
%\apjref White, S.D.M. and Audouze, J. 1983;\mnras;203;603
\apjref  Wyse, R.F.G. and Gilmore, G. 1988;\aj;95;1404
%\apjref  Wyse, R.F.G. and Gilmore, G.;1989;{Comments on Ap.};8;135
\apjref Wyse, R.F.G. and Gilmore, G. 1992;\aj;104;144
\pp Wyse, R.F.G. and Gilmore, G. 1994 (in preparation)

\vfill\eject

\centerline  {FIGURE CAPTIONS}
\parindent=0pt

\pp
Figure 1(a) : B$-$V,  V color-magnitude diagram for stars in the
metallicity sample in the South Galactic Pole field.

\pp
Figure 1(b) : V$-$I, V color-magnitude diagram for stars in the
metallicity sample in the South Galactic Pole field.

\pp
Figure 2 : B$-$V, V$-$I two-color diagram for  stars in the
metallicity sample in the South Galactic Pole field, together with the
ridge line for E-region standard stars from Reid and Gilmore (1982).

\pp Figure 3 : B$-$V (circles) and V$-$I (squares) for one solar mass
models, at fixed effective temperature, and age of many Gyr,
from VandenBerg and Bell (1985; open symbols) and VandenBerg
(1985; filled symbols) as a function of metallicity.

\pp
Figure 4 : B$-$V, V color-magnitude diagram  for all stars in the
chosen magnitude range in the
SGP fields.

\pp
Figure 5(a) : B$-$V, V$-$I two-color diagram for  stars in the
metallicity sample in the South Galactic Pole field, with different
ranges of metallicity plotted separately.

\pp
Figure 5(b) : B$-$V, V$-$I two-color diagram for  the brighter stars
(V $\leq 17.30$) in the
metallicity sample in the South Galactic Pole field, with different
ranges of metallicity plotted separately.

\pp Figure 6 :
Scatter plot of iron abundance versus B$-$V color
for the brighter stars in the SGP, together with the 12 Gyr turnoff
points from VandenBerg and Bell (1985; crosses) and from VandenBerg
(1985; asterisks).

\pp Figure 7 : Comparison of the iron abundance histograms for the
`raw' bright SGP sample (dashed lines) and for the color-corrected and turnoff
corrected bright sample (solid  lines).

\pp Figure 8 : Scatter plot of iron abundance versus distance for the
color-corrected and turnoff-corrected bright SGP sample.

\pp Figure 9 : Scatter plot of B$-$V versus distance for the SGP sample
of Figure 8, together with the  distance limits corresponding to the
bright
(dashed lines) and
faint (solid lines)
magnitude selection, for metallicities from left to right of
$-2$dex to solar, in
steps of 0.5dex.

\pp Figure 10 : Histograms of the volume-complete iron abundance
distributions for F/G dwarfs in the SGP at 1500pc above the plane, derived by
(i)
assuming all stars follow the thick disk density profile, indicated by
the solid lines, and (ii) assuming the most metal-poor bins are
entirely stellar halo stars, and the most metal-rich are thin disk
stars, as described in the text (dashed lines).

\pp Figure 11(a) : B$-$V, V color-magnitude diagram for stars in the
metallicity sample in F117.

\pp
Figure 11(b) : V$-$I, V color-magnitude diagram for stars in the
metallicity sample in F117.

\pp
Figure 12 : B$-$V, V color-magnitude diagram  for stars in the
relevant magnitude range in
F117 fibre fields.

\pp
Figure 13 : B$-$V, V$-$I two-color diagram for  stars in the
metallicity sample in F117.

\pp
Figure 14 : B$-$V, V$-$I two-color diagram for  stars in the
metallicity sample in F117,  with different
ranges of metallicity plotted separately.

\pp
Figure 15 : Scatter plot of iron abundance versus B$-$V color
for the  stars in F117, together with the 12 Gyr turnoff
points from VandenBerg and Bell (1985; crosses) and from VandenBerg
(1985; asterisks).

\pp Figure 16 :
 Scatter plot of iron abundance versus distance for the
F117 stars.

\pp Figure 17 : Scatter plot of B$-$V versus distance for the F117
stars,
together with the distance limits corresponding to the bright (dashed
lines) and
faint (solid lines) magnitude selection, for metallicities from left to
right of  $-2$dex to solar, in
steps of 0.5dex.

\pp Figure 18 : Histograms of the volume-complete iron abundance
distributions for F/G dwarfs in F117 at 1000pc above the plane, derived by (i)
assuming all stars follow the thick disk density profile, indicated by
the solid lines, and (ii) assuming the most metal-poor bins are
entirely stellar halo stars, and the most metal-rich are thin disk
stars, as described in the text (dashed lines).
Note that those bins with [Fe/H]$\simlt -1$dex should formally be treated
as lower limits.

\pp Figure 19 : Comparison of the F117 $z=1000$pc iron-abundance
distribution (solid histograms) and the SGP $z=1500$pc distribution
(dashed histograms).  The left panel shows the derived distributions
under the assumption that all stars follow the thick-disk density
profile, while the right panel includes thin disk and stellar halo
contributions.

\vfill\eject
\baselineskip=20pt
\parskip=3pt plus 1pt minus 1pt  %it is zero plus 1pt in plain.tex
  \tabskip=.2em plus .5em minus .2em
  \def \space{ \noalign{ \vskip4pt \hrule height 1pt \vskip4pt} }
  $$ \vbox{ \halign to \hsize { #\hfil & \hfil#\hfil & \hfil#\hfil &
  \hfil#\hfil & \hfil#\hfil & \hfil#\hfil \cr
 \multispan{6} \hfil {TABLE 1 : {Relative error contributions in [Fe/H]
determination}} \hfil \cr
 \noalign{ \vskip 12pt }
 \noalign{ \hrule height 1pt \vskip 1pt \hrule height 1pt }
 \noalign{ \vskip 8pt }
T$_{eff}$\phantom{spa} & Error in T$_{eff}$ & $\rm \Delta \left[{Fe
\over H}\right]_{[{Fe\over H}]=-0.5}$ & $\rm \Delta \left[{Fe
\over H}\right]_{[{Fe\over H}]=-1.5}$ & $\rm \Delta \left[{Fe
\over H}\right]_{[{Fe\over H}]=-0.5}$ & $\rm \Delta \left[{Fe
\over H}\right]_{[{Fe\over H}]=-1.5}$ \cr
 & $\sigma _ {\rm V-I} = 0.05 $ & $\sigma _ {\rm V-I} = 0.05 $ &
$\sigma _ {\rm V-I} = 0.05 $ &
S/N = 15/\AA & S/N = 15/\AA\cr
 \noalign{ \vskip 8pt\hrule height 1pt \vskip 8pt }
5000K&140K&0.13&0.16&0.12&0.15 \cr
5500K&170K&0.13&0.16& \cr
6000K&210K&0.14&0.17&0.09&0.25 \cr
 \noalign{ \vskip 8pt\hrule height 1pt
  \vskip 1pt \hrule height 1pt \vskip 8pt }
}}$$
\vfill\eject
\parskip=3pt plus 1pt minus 1pt  %it is zero plus 1pt in plain.tex
  \tabskip=.2em plus .5em minus .2em
  \def \space{ \noalign{ \vskip4pt \hrule height 1pt \vskip4pt} }
  $$ \vbox{ \halign to \hsize { \hfil#\hfil & \hfil#\hfil & \hfil#\hfil
& \hfil#\hfil &
  \hfil#\hfil & \hfil#\hfil & \hfil# & \hfil#\hfil \cr
 \multispan{8} \hfil {TABLE 2 : {Observed and Derived Data for SGP
stars}}
\hfil \cr
 \noalign{ \vskip 12pt }
 \noalign{ \hrule height 1pt \vskip 1pt \hrule height 1pt }
 \noalign{ \vskip 8pt }
      ra (1950.0) &   dec (1950.0)  &    V &   B--V &  V--I & [Fe/H]&
dist(pc)&  Teff
\cr
 \noalign{ \vskip 8pt\hrule height 1pt
  \vskip 8pt }
 \phantom{0}0  47  39.40 & $-$29   13 31.1  & 16.47  & 0.80  & 0.73 &
\phantom{$-$}0.16 & 1478 & 5602 \cr
 \phantom{0}0  47  51.26 & $-$29   10 14.5  & 16.67  & 0.62  & 0.93 & $-$0.29 &
2247 & 4961 \cr
 \phantom{0}0  48  \phantom{0}0.56 & $-$29    06   \phantom{0}5.5  &
17.24  & 0.62  & 0.84 & $-$0.78  & 2273 & 5231 \cr
 \phantom{0}0  48    \phantom{0}5.89 & $-$28   59 59.9  & 16.84  & 0.68
& 1.00 & $-$0.90 & 1553 & 4770 \cr
 \phantom{0}0  48  11.66 & $-$28   59   \phantom{0}4.5  & 15.90  & 0.50  &
0.63 & $-$0.84 & 1576 & 5989 \cr
 \phantom{0}0  48  17.96 & $-$29   41   \phantom{0}2.1  & 17.33  & 0.63  &
0.96 & $-$0.92 & 2165 & 4877 \cr
 \phantom{0}0  48  18.35 & $-$29    \phantom{0}0 56.9  & 16.38  & 0.64 &
0.74  & $-$0.25 & 1907 & 5566 \cr
 \phantom{0}0  48  20.79 & $-$29   16 52.8  & 16.97  & 0.64  & 0.80 & $-$0.30 &
2441 & 5360 \cr
 \phantom{0}0  48  25.88 & $-$29   21 47.8  & 16.23  & 0.89  & 0.98 & $-$0.17
&
\phantom{1}978 & 4823 \cr
 \phantom{0}0  48  26.71 & $-$29   45 11.5  & 17.88  & 0.52  & 0.74 & $-$1.19 &
3164 & 5566 \cr
 \phantom{0}0  48  30.53 & $-$29   48 44.9  & 16.29  & 0.49  & 0.65 & $-$0.72 &
2058 & 5908 \cr
 \phantom{0}0  48  34.00 & $-$29   52 56.7  & 17.18  & 0.85  & 0.95 &
\phantom{$-$}0.13 &
1800 & 4905 \cr
 \phantom{0}0  48  35.40 & $-$29   14 14.5  & 17.31  & 0.63  & 0.88 & $-$0.56 &
2561 & 5107 \cr
  \phantom{0}0  48  36.67 & $-$29    \phantom{0}7 22.0  & 15.97  & 0.70 &
0.77  & \phantom{$-$}0.03 & 1512 & 5461 \cr
  \phantom{0}0  48  41.27 & $-$29   47 56.8  & 16.84  & 0.87  &  1.00 & $-$0.34
&
1270 & 4770 \cr
  \phantom{0}0  48  49.64 & $-$29   36 16.9  & 17.92  & 0.46 & 0.75 & $-$0.67 &
4813 & 5531 \cr
  \phantom{0}0  48  54.26 & $-$30    \phantom{0}4 57.0  & 17.25  & 0.78 &
1.02  & $-$0.75 & 1588 & 4718 \cr
  \phantom{0}0  49    \phantom{0}2.00 & $-$29   30 14.7  & 16.26  & 0.54  &
0.73 & $-$1.09 & 1501
& 5602 \cr
  \phantom{0}0  49    \phantom{0}3.07 & $-$29   15 26.9  & 16.29  & 0.74
& 0.81 & $-$0.50 & 1262 & 5327 \cr
  \phantom{0}0  49    \phantom{0}3.95 & $-$29   35 59.5  & 17.49  & 0.55
& 0.78 & $-$0.98 &
2723 & 5427 \cr
  \phantom{0}0  49    \phantom{0}6.70 & $-$29   18 33.4  & 16.35  & 0.64  &
0.74 & $-$0.22 &
1861 & 5461 \cr
  \phantom{0}0  49    \phantom{0}6.83 & $-$30    \phantom{0}4 57.8  & 16.48  &
0.86  &
0.90 & $-$0.29 & 1127 & 5048 \cr
  \phantom{0}0  49  12.50 & $-$30    \phantom{0}2 46.9  & 15.79  & 0.65  &
0.68 & $-$0.60 & 1187 & 5789 \cr
    \noalign{ \vskip 8pt\hrule height 1pt }
}}$$
\vfill\eject

  $$ \vbox{ \halign to \hsize { \hfil#\hfil & \hfil#\hfil & \hfil#\hfil &
  \hfil#\hfil & \hfil#\hfil & \hfil#\hfil & \hfil# & \hfil#\hfil \cr
 \multispan{8} \hfil {TABLE 2 (continued) }
\hfil \cr
 \noalign{ \vskip 12pt }
 \noalign{ \hrule height 1pt }
 \noalign{ \vskip 8pt }
      ra (1950.0) &   dec (1950.0)  &    V &   B--V & V--I & [Fe/H]&  dist(pc)&
 Teff
\cr
 \noalign{ \vskip 8pt\hrule height 1pt
  \vskip 8pt }
  \phantom{0}0  49  14.60 & $-$30    \phantom{0}2   \phantom{0}6.5  & 16.82  &
0.76 & 0.92 & $-$0.87 & 1295
& 4990 \cr
  \phantom{0}0  49  18.01 & $-$30    \phantom{0}5 53.6  & 16.30  & 0.71  &
0.79 &    $-$0.11 &
1639 & 5393 \cr
  \phantom{0}0  49  18.32 & $-$29   34 45.5  & 16.55  & 0.77  &  0.87 & $-$0.20
& 1519 & 5138 \cr
  \phantom{0}0  49  18.45 & $-$30    \phantom{0}9 47.0  & 17.68  & 0.95 &
0.97   & $-$0.55 &
1405 & 4850 \cr
  \phantom{0}0  49  21.46 & $-$29   54 10.5  & 16.87  & 0.63  & 0.86 & $-$0.35
& 2330 & 5168 \cr
  \phantom{0}0  49  23.26 & $-$29   53 34.9  & 15.90  & 0.85  & 1.04 & $-$1.88
&  \phantom{1}494
& 4667 \cr
  \phantom{0}0  49  24.78 & $-$29   57 22.3  & 16.31  & 0.83  & 0.89 &
\phantom{$-$}0.12 &
1270 & 5077 \cr
  \phantom{0}0  49  30.85 & $-$29   24 51.7  & 16.78  & 0.70  & 0.80  &
\phantom{$-$}0.21 & 2210 & 5360 \cr
  \phantom{0}0  49  41.81 & $-$29   22 22.7  & 16.03  & 0.78 &   0.83 & $-$0.12
&
1205 & 5262 \cr
  \phantom{0}0  49  43.22 & $-$29   54 48.9  & 17.27  & 0.57  & 0.84 & $-$0.96
& 2373 & 5231 \cr
  \phantom{0}0  49  47.37 & $-$29   59 55.5  & 15.99  & 0.57  &  0.69 & $-$0.63
&
1554 & 5751 \cr
  \phantom{0}0  49  52.71 & $-$29   58 19.5  & 16.08  & 0.69  &  0.79 & $-$0.56
&
1257 & 5393 \cr
\phantom{0}0   50    \phantom{0}2.47 & $-$29   52 35.9  & 16.86  & 0.63  & 0.78
& $-$0.66 & 1978 & 5427 \cr
  \phantom{0}0  50    \phantom{0}3.56 & $-$29   57   \phantom{0}2.5  &
17.62  & 0.98  & 0.91 &   $-$0.30 & 1404 & 5019 \cr
  \phantom{0}0  50  11.10 & $-$29   57 58.0  & 17.69  & 0.30  &
0.70 &  $-$1.80 & 3645 & 5713 \cr
  \phantom{0}0  51   \phantom{0}4.66 & $-$29    \phantom{0}7 58.0   & 16.84  &
0.58
&0.89 & $-$0.38 & 2562 & 5077 \cr
  \phantom{0}0  51  14.84 & $-$28   58 46.8   & 16.97  & 0.82  & 0.87 &
$-$0.02 & 1735 & 5138 \cr
  \phantom{0}0  51  17.44 & $-$29    \phantom{0}8 27.0   & 16.78  & 0.63
&0.82  & $-$0.90 & 1696 & 5294 \cr
  \phantom{0}0  51  20.06 & $-$29    \phantom{0}6 30.2   & 17.51  & 0.76
&0.93 &$-$0.53 &
2079 & 4961 \cr
  \phantom{0}0  51  23.65 & $-$29    \phantom{0}4 55.1   & 17.46  & 0.59 &
0.81 & $-$1.68 & 1852 & 5327 \cr
  \phantom{0}0  51  32.88 & $-$28   56 44.7   & 16.90  & 0.67 & 0.82 & $-$0.73
& 1770 & 5294 \cr
  \phantom{0}0  51  40.45 & $-$29    \phantom{0}0 25.7   & 16.95  & 0.56 &
0.71  & $-$0.78 &
2291 & 5676 \cr
  \phantom{0}0  51  41.08 & $-$29   14 40.8  & 17.26  & 0.57 & 0.81  & $-$1.21
& 2108 & 5327 \cr
    \noalign{ \vskip 8pt\hrule height 1pt }
}}$$
\vfill\eject

  $$ \vbox{ \halign to \hsize { \hfil#\hfil & \hfil#\hfil & \hfil#\hfil &
  \hfil#\hfil & \hfil#\hfil & \hfil#\hfil & \hfil# & \hfil#\hfil \cr
 \multispan{8} \hfil {TABLE 2 (continued) }
\hfil \cr
 \noalign{ \vskip 12pt }
 \noalign{ \hrule height 1pt }
 \noalign{ \vskip 8pt }
      ra (1950.0) &   dec (1950.0)  &    V &   B--V & V--I & [Fe/H]&  dist(pc)&
 Teff
\cr
 \noalign{ \vskip 8pt\hrule height 1pt
  \vskip 8pt }
  \phantom{0}0  51  35.12 & $-$28   53 16.6   & 17.60  & 0.34 & 1.01 & $-$1.28
& 3982 & 4744 \cr
  \phantom{0}0  51  52.94 & $-$28   50 07.5   & 17.78  & 0.84 & 0.98 & $-$0.42
&
2036 & 4823 \cr
  \phantom{0}0  52  27.20 & $-$29    \phantom{0}6 39.0   & 17.37  & 0.49 &
0.85  & $-$1.28 & 2570 & 5199 \cr
  \phantom{0}0  52  29.58 & $-$29   17 21.5   & 16.95  & 0.86 & 0.91   &
$-$0.41 & 1328 & 5138 \cr
  \phantom{0}0  52  35.17 & $-$28   53 31.2   & 17.35  & 0.78 & 0.97 & $-$0.44
& 1918 & 5019 \cr
  \phantom{0}0  52  48.90 & $-$28   59 28.6   & 17.58  & 0.89 & 0.99 & $-$1.13
&
1231 & 4850 \cr
  \phantom{0}0  52  57.14 & $-$29   10 54.4   & 17.42  & 1.00  & 0.87 & $-$0.12
& 1301 & 4796 \cr
  \phantom{0}0  53   \phantom{0}6.62 & $-$28   17 48.7  & 17.30  & 0.73
&0.96 & $-$0.36 &
2204 & 4877 \cr
  \phantom{0}0  53  37.02 & $-$29    \phantom{0}7 44.1   & 17.01  & 0.96 &
0.76 & $-$0.21 &
1155 & 5496 \cr
  \phantom{0}0  53  47.39 & $-$28   26 21.5  & 16.75  & 0.97 & 0.96 & $-$0.21 &
 \phantom{1}999 & 4877 \cr
  \phantom{0}0 57   \phantom{0}4.21 & $-$29 59 30.6  & 16.76 &
0.54 & 0.77  & $-$0.82 & 2155 & 5461 \cr
  \phantom{0}0 57   \phantom{0}8.76 & $-$30   \phantom{0}5 30.7  & 16.74 & 0.83
& 0.82 & \phantom{$-$}0.18  &
1548 & 5294 \cr
  \phantom{0}0 57 10.10 & $-$28 58   \phantom{0}3.3  & 17.10 & 0.54   &
0.68 &  $-$0.69 & 2697 & 5789 \cr
  \phantom{0}0 57 14.53 & $-$29   \phantom{0}4 51.5  & 17.05 & 0.56&0.74    &
$-$0.49 & 2795
& 5566 \cr
  \phantom{0}0 57 29.50 & $-$28 50 45.6  & 17.11 & 0.40  & 0.90 & $-$0.70 &
3759 & 5048 \cr
  \phantom{0}0 57 30.39 & $-$29 49 17.0  & 16.66 & 0.61  & 0.77  & $-$0.94 &
1650 & 5461 \cr
  \phantom{0}0 57 30.54 & $-$29   \phantom{0}7 28.7  & 16.90 &
0.58 & 0.74  & $-$0.31 &
2733 & 5566 \cr
  \phantom{0}0 57 32.88 & $-$27 55   \phantom{0}0.5  & 16.23 &
0.63 & 0.66 & $-$0.13 & 1931 & 5868 \cr
  \phantom{0}0 57 37.09 & $-$29   \phantom{0}7 50.0  & 17.02 &
0.61 & 0.89  & $-$1.51 & 1536 & 5077 \cr
  \phantom{0}0 57 38.02 & $-$30   \phantom{0}8 43.6  & 16.68 &
0.89  & 0.98 & $-$0.33 &
1127 & 4823 \cr
  \phantom{0}0 57 43.60 & $-$28   \phantom{0}3 44.3  & 16.77 &
0.78  & 0.84 & $-$0.11 &
1701 & 5231 \cr
  \phantom{0}0 57 43.88 & $-$29   \phantom{0}4 33.0  & 17.26 &
0.75  & 0.77 & $-$0.06 &
2345 & 5461 \cr
  \phantom{0}0 57 45.35 & $-$30   \phantom{0}9  2.1  & 16.22 &
0.48  & 0.72   & $-$1.45 &
1441 & 5639 \cr
    \noalign{ \vskip 8pt\hrule height 1pt }
}}$$
\vfill\eject

  $$ \vbox{ \halign to \hsize { \hfil#\hfil & \hfil#\hfil & \hfil#\hfil &
  \hfil#\hfil & \hfil#\hfil & \hfil#\hfil & \hfil# & \hfil#\hfil \cr
 \multispan{8} \hfil {TABLE 2 (continued) }
\hfil \cr
 \noalign{ \vskip 12pt }
 \noalign{ \hrule height 1pt }
 \noalign{ \vskip 8pt }
      ra (1950.0) &   dec (1950.0)  &    V &   B--V & V--I & [Fe/H]&  dist(pc)&
 Teff
\cr
 \noalign{ \vskip 8pt\hrule height 1pt
  \vskip 8pt }
  \phantom{0}0 57 45.68 & $-$28 37 16.3  & 16.26 & 0.84  & 0.85 &
\phantom{$-$}0.09 & 1210 & 5199 \cr
  \phantom{0}0 57 45.72 & $-$29 56 28.5  & 17.26 & 0.65  &  0.81 & $-$0.07 &
3022 & 5327 \cr
  \phantom{0}0 57 46.24 & $-$29 10 15.2  & 17.66 & 0.75  & 0.80 & $-$0.10 &
2778 & 5360 \cr
  \phantom{0}0 57 48.92 & $-$30   \phantom{0}1   \phantom{0}9.5  & 16.92 & 0.77
&0.90&   $-$0.06 & 1904
& 5048 \cr
  \phantom{0}0 57 52.53 & $-$29 44 38.6  & 16.78 & 0.52  & 0.83 & $-$0.76 &
2352 & 5262 \cr
  \phantom{0}0 57 55.16 & $-$29 55 14.8  & 16.68 & 0.78  & 0.92 & $-$0.50 &
1370 & 4990 \cr
  \phantom{0}0 57 55.31 & $-$29 43 23.2  & 16.02 & 0.63  & 0.81 & $-$0.88 &
1207 & 5327 \cr
  \phantom{0}0 57 57.45 & $-$28 48 45.2  & 16.55 & 0.55  & 0.69 & $-$0.43 &
2350 & 5751 \cr
  \phantom{0}0 58   \phantom{0}0.10 & $-$28 15 55.0  & 17.22 & 0.66  & 0.80 &
$-$0.95 & 1894 & 5360 \cr
  \phantom{0}0 58   \phantom{0}2.18 & $-$28   \phantom{0}3 30.2  & 17.25 & 0.74
 &
0.80& $-$0.21 &
2253 & 5360 \cr
  \phantom{0}0 58   \phantom{0}2.37 & $-$29 58 20.3  & 16.59 &
0.86 & 0.96  & $-$0.34 & 1160 & 4877 \cr
  \phantom{0}0 58   \phantom{0}3.16 & $-$28 55 32.6  & 17.03 & 0.51  &
0.63 & $-$0.54 & 3044 & 5989 \cr
  \phantom{0}0 58   \phantom{0}9.62 & $-$29   \phantom{0}1 49.3  & 17.42 & 0.50
 &
0.68 & $-$0.67 & 3475 & 5789 \cr
  \phantom{0}0 58 10.50 & $-$28 49 36.5  & 16.28 & 0.62  & 0.82 & $-$0.46 &
1720 & 5294 \cr
  \phantom{0}0 58 11.42 & $-$28 14 50.7  & 16.28 & 0.54  & 0.72 & $-$0.60 &
1940 & 5639 \cr
  \phantom{0}0 58 13.39 & $-$28 22  3.5  & 17.44 & 0.78  & 0.87 & $-$0.50 &
1944 & 5138 \cr
  \phantom{0}0 58 14.45 & $-$29   \phantom{0}8 47.7  & 16.40 &
0.82  & 0.93 & $-$0.20 &
1249 & 4961 \cr
  \phantom{0}0 58 14.98 & $-$28 59 28.4  & 17.22 & 0.45  & 0.61 & $-$1.24 &
2668 & 6073 \cr
  \phantom{0}0 58 19.20 & $-$29 52 13.0  & 16.39 & 0.54  & 0.73 & $-$0.21 &
2517 & 5602 \cr
  \phantom{0}0 58 19.25 & $-$28 51   \phantom{0}3.8  & 16.76 &
0.53 & 0.67  & $-$0.45 & 2690 & 5828 \cr
  \phantom{0}0 58 19.46 & $-$28   \phantom{0}7 22.0  & 16.38 &
0.68 & 0.81 & $-$0.43 &
1578 & 5327 \cr
  \phantom{0}0 58 19.58 & $-$27 53 46.0  & 16.45 & 0.82  & 0.97 & $-$0.71 &
1017 & 4850 \cr
  \phantom{0}0 58 23.77 & $-$28 17 44.2  & 16.96 & 0.77  & 0.82 & $-$0.31 &
1746 & 5294 \cr
 \noalign{ \vskip 8pt\hrule height 1pt
  \vskip 8pt }
}}$$
\vfill\eject

  $$ \vbox{ \halign to \hsize { \hfil#\hfil & \hfil#\hfil & \hfil#\hfil &
  \hfil#\hfil & \hfil#\hfil & \hfil#\hfil & \hfil# & \hfil#\hfil \cr
 \multispan{8} \hfil {TABLE 2 (continued)}
\hfil \cr
 \noalign{ \vskip 12pt }
 \noalign{ \hrule height 1pt }
 \noalign{ \vskip 8pt }
      ra (1950.0) &   dec (1950.0)  &    V &   B--V & V--I & [Fe/H]&  dist(pc)&
 Teff
\cr
 \noalign{ \vskip 8pt\hrule height 1pt
  \vskip 8pt }
  \phantom{0}0 58 31.73 & $-$27 55 35.7  & 16.82 & 0.82  & 0.86 & $-$1.56 &
\phantom{1}877 & 5168 \cr
  \phantom{0}0 58 32.19 & $-$28 48 34.8  & 17.62 & 0.50  & 0.83   & $-$0.94 &
3307 & 5262 \cr
  \phantom{0}0 58 37.01 & $-$28 25   \phantom{0}0.0  & 16.22 &
0.62  & 0.86 & $-$0.73 &
1457 & 5168 \cr
  \phantom{0}0 58 37.35 & $-$29 41 27.8  & 16.12 & 0.68  & 0.79 & $-$0.38 &
1435 & 5393 \cr
  \phantom{0}0 58 39.88 & $-$28   \phantom{0}8 38.4  & 17.90 &
0.57  & 0.82 & $-$1.27 & 2760 & 5294 \cr
  \phantom{0}0 58 40.18 & $-$29 50 31.6  & 16.32 & 0.59  & 0.68 & $-$0.54 &
1807 & 5789 \cr
  \phantom{0}0 58 40.20 & $-$30   \phantom{0}0 56.8  & 16.10 & 0.73  & 0.77 &
$-$0.05 & 1452 & 5461 \cr
  \phantom{0}0 58 40.86 & $-$29 59 34.7  & 17.70 & 0.67  & 0.82 & $-$1.05 &
2209 & 5294 \cr
  \phantom{0}0 58 41.22 & $-$27 49 37.8  & 17.40 & 0.70  & 0.81 & $-$0.47 &
2355 & 5327 \cr
  \phantom{0}0 58 54.84 & $-$28   \phantom{0}2 17.9  & 16.61 &
0.54 & 0.70 & $-$0.94 & 1894 & 5713 \cr
  \phantom{0}0 58 55.43 & $-$27 51 25.3  & 17.34 & 0.80  & 0.86 & $-$0.18 &
2043 & 5168 \cr
  \phantom{0}0 58 57.74 & $-$29 55 11.8  & 17.35 & 0.63  & 0.93 & $-$0.64 &
2504 & 4961 \cr
  \phantom{0}0 59   \phantom{0}2.18 & $-$28 10 31.6  & 16.20 &
0.76 & 0.71 & \phantom{$-$}0.25 & 1448 & 5676 \cr
  \phantom{0}0 59   \phantom{0}5.41 & $-$28 57 13.5  & 16.39 &
0.86  & 0.98 & $-$0.21 & 1118 & 4823 \cr
  \phantom{0}0 59   \phantom{0}7.46 & $-$28   \phantom{0}2 51.1  & 17.44 & 0.58
 &
0.97 & $-$1.26 & 2193 & 4850 \cr
  \phantom{0}0 59   \phantom{0}7.74 & $-$28   \phantom{0}0   \phantom{0}4.8  &
16.50 & 0.57  &
0.60 & $-$0.38 &
2246 & 6116 \cr
  \phantom{0}0 59 15.75 & $-$28 12 38.5  & 17.24 & 0.62 & 0.90 &  $-$1.09 &
1967 & 5048 \cr
  \phantom{0}0 59 24.51 & $-$28   \phantom{0}4 29.9  & 17.11 & 0.53 &  0.66 &
$-$0.68 & 2789 & 5868 \cr
  \phantom{0}0 59 25.38 & $-$27 54 28.3  & 16.73 & 0.89 & 0.92 &   $-$0.29 &
1173 & 4990 \cr
  \phantom{0}0 59 29.09 & $-$28 13 17.7  & 16.55 & 0.64  & 0.63 &
\phantom{$-$}0.00 & 2290 & 5989 \cr
  \phantom{0}0 59 36.01 & $-$27 41   \phantom{0}1.9  & 16.49 &
0.76 & 0.91  & $-$0.53 &
1300 & 5019 \cr
  \phantom{0}0 59 44.38 & $-$27 53 55.0  & 16.91 & 0.85 & 0.87  & $-$0.36 &
1366 & 5138 \cr
  \phantom{0}0 59 53.23 & $-$27 39 30.9  & 16.75 & 0.64 & 0.76 & $-$0.50 &
1992 & 5496 \cr
 \noalign{ \vskip 8pt\hrule height 1pt
  \vskip 8pt }
}}$$
\vfill\eject
  $$ \vbox{ \halign to \hsize { \hfil#\hfil & \hfil#\hfil & \hfil#\hfil &
  \hfil#\hfil & \hfil#\hfil & \hfil#\hfil & \hfil# & \hfil#\hfil \cr
 \multispan{8} \hfil {TABLE 2 (continued)}
\hfil \cr
 \noalign{ \vskip 12pt }
 \noalign{ \hrule height 1pt }
 \noalign{ \vskip 8pt }
      ra (1950.0) &   dec (1950.0)  &    V &   B--V &  V--I & [Fe/H]&
dist(pc)&  Teff
\cr
 \noalign{ \vskip 8pt\hrule height 1pt
  \vskip 8pt }
  \phantom{0}0 59 55.65 & $-$27 44 39.0  & 16.63 & 0.50 & 0.71  & $-$0.46 &
2713 & 5676 \cr
  \phantom{0}0 59 56.84 & $-$27 41 27.1  & 16.29 & 0.59 & 0.71  & $-$0.83 &
1536 & 5676 \cr
   1   \phantom{0}0   \phantom{0}1.83 & $-$27 45 50.9  & 16.57 &
0.61 & 0.72  & $-$1.27 & 1369 & 5639 \cr
   1   \phantom{0}0   \phantom{0}3.58 & $-$27 31   \phantom{0}4.6  & 16.81 &
0.86  &
1.02 & $-$0.67 &
1111 & 4718 \cr
   1   \phantom{0}0   \phantom{0}3.72 & $-$27 50 11.0  & 16.55 & 0.76  &
0.82 & $-$0.22 & 1544 & 5294 \cr
   1   \phantom{0}0 12.17 & $-$27 37 47.2  & 17.61 & 0.49  & 0.75
&  $-$1.68 & 2454 & 5531 \cr
   1   \phantom{0}0 32.51 & $-$27 53 56.7  & 16.64 & 0.71 & 0.91 & $-$1.13 &
1197 & 5019 \cr
   1 \phantom{0}0 36.20 & $-$27 42   \phantom{0}0.1  & 17.08 &
0.72  & 0.86 &   $-$0.52 &
1889 & 5168 \cr
   1 \phantom{0}0 42.13 & $-$27 40   \phantom{0}1.4  & 17.48 &
0.80  & 0.82 &  $-$0.40 & 1975 & 5294 \cr
   1 \phantom{0}0 44.02 & $-$27 53 52.6  & 16.69 & 0.78  & 0.90 & $-$0.82 &
1190 & 5048 \cr
   1 \phantom{0}0 44.44 & $-$27 48   \phantom{0}0.0  & 17.48 &
0.82  & 0.82 &   $-$0.49 & 1804 & 5294 \cr
   1 \phantom{0}0 46.68 & $-$27 42 48.5  & 17.68 & 0.73  & 0.97 & $-$0.94 &
2001 & 4850 \cr
   1 \phantom{0}0 48.19 & $-$27 45 16.8  & 16.99 & 0.58  & 0.69 &   $-$0.49 &
2589 & 5751 \cr
   1 \phantom{0}0 52.27 & $-$27 55 15.4  & 16.56 & 0.51 & 0.68 & $-$0.78 & 2153
& 5789 \cr
   1 \phantom{0}0 55.22 & $-$27 33   \phantom{0}2.8  & 16.35 &
0.68 & 0.64 & $-$0.40 & 1579 & 5948 \cr
   1 \phantom{0}0 57.80 & $-$27 23   \phantom{0}5.7  & 16.68 & 0.45  & 0.98 &
$-$2.43 &
1421 & 4823 \cr
   1   \phantom{0}1 15.29 & $-$27 46 26.5  & 16.59 & 0.87  & 0.90
&  $-$0.16 &
1220 & 5048 \cr
   1   \phantom{0}1 19.30 & $-$27 44 19.3  & 17.38 & 0.62  &  0.75   & $-$0.54
&
2738 & 5531 \cr
 \noalign{ \vskip 8pt\hrule height 1pt \vskip 1pt \hrule height 1pt
  \vskip 8pt }
}}$$

\vfill\eject

\parskip=3pt plus 1pt minus 1pt  %it is zero plus 1pt in plain.tex
  \tabskip=.2em plus .5em minus .2em
  \def \space{ \noalign{ \vskip4pt \hrule height 1pt \vskip4pt} }
  $$ \vbox{ \halign to \hsize { \hfil#\hfil & \hfil# & \hfil# &
  \hfil# & \hfil# & \hfil# & \hfil# \cr
 \multispan{7} \hfil {TABLE  3 : {SGP Iron Abundance Distribution }}\hfil \cr
 \noalign{ \vskip 12pt }
 \noalign{ \hrule height 1pt \vskip 1pt \hrule height 1pt }
 \noalign{ \vskip 8pt }
Central [Fe/H] & 1  & 2 & 3 & 4 & 5 & 6 \cr
 \noalign{ \vskip 8pt \hrule height 1pt }
 \noalign{ \vskip 8pt }
+0.3 & 2 &  2 & 2290 & 1960 & 2 & 6 \cr
+0.1 & 7 & 7 & 2290 & 1960 & 8 & 22 \cr
$-$0.1 & 13 & 13 & 2180 & 2150 &  15 & 17 \cr
$-$0.3 & 21 & 21 & 1995 & 1970 & 28 & 29 \cr
$-$0.5 & 18 & 19 & 1810 & 1810 & 29 & 29 \cr
$-$0.7 & 15 & 15 & 1640 & 1640 & 27 & 27 \cr
$-$0.9 & 12 & 13 & 1500 & 1500 & 28 & 28 \cr
$-$1.1 & 3 & 3 & 1370 & 1430 & 8 & 9 \cr
$-$1.3 & 3 & 4 & 1260 & 1310 & 12 & 14 \cr
$-$1.5 & 3 & 4 & 1170 & 1215 & 14 & 19  \cr
$-$1.7 & 0 & 0 & 1095  & 1135 & 0 & 0 \cr
$-$1.9 & 1 & 1 & 1035 & 1070 & 5 & 7 \cr
$-$2.1 & 0 & 0 & 985 & 1015 & 0 & 0  \cr
$-$2.3 & 0 & 0 & 945 & 970 & 0 & 0 \cr
$-$2.5 & 1 & 2 & 910 &   940 & 12 & 19 \cr
 \noalign{ \vskip 8pt\hrule height 1pt \vskip 1pt \hrule height 1pt
  \vskip 8pt }
}}$$
\baselineskip=12pt

\centerline {Notes to TABLE 3}

Column number 1 contains the `raw' observed metallicity
distribution for the brighter stars with V$ < 17.30$ and
B$-$V $< 0.9$; column
number 2  the B$-$V bias-corrected
and turnoff-corrected sample; column number 3
the characteristic distance (in pc) of each metallicity bin if all stars are
assumed to follow the thick disk density profile; column number 5
the number of stars per unit volume at a fiducial
distance of 1500pc under this assumed
density law; column number 4  the characteristic distances if
stellar halo and thin disk contributions are introduced, as explained
in the text; column number 6  the number of stars
per unit volume at 1500pc under these assumed density laws.
\vfil\eject
\baselineskip=20pt
  $$ \vbox{ \halign to \hsize { \hfil#\hfil & \hfil#\hfil & \hfil#\hfil &
  \hfil#\hfil & \hfil#\hfil & \hfil#\hfil & \hfil#  & \hfil#\hfil \cr
 \multispan{8} \hfil {TABLE 4 {Observed and Derived Data for F117
stars}}
\hfil \cr
 \noalign{ \vskip 12pt }
 \noalign{ \hrule height 1pt \vskip 1pt \hrule height 1pt }
 \noalign{ \vskip 8pt }
  ra (1950.0) & dec (1950.0)  &  V & B--V &  V--I & [Fe/H]&  dist(pc)&  Teff
\cr
 \noalign{ \vskip 8pt\hrule height 1pt
  \vskip 8pt }
 3 40  05.00  & $-$59 17 24.3 & 16.89  & 0.64 &  0.75 & $-$0.40 & 2236 & 5478
\cr
 3 40 10.70  & $-$59 17 32.3 & 17.27  & 0.63 & 0.71 & $-$0.77 & 2262 & 5620 \cr
 3 40 24.33  & $-$59 31 45.4 & 17.28  & 0.75 & 0.97 & $-$0.97 & 1569 & 4810 \cr
 3 40 27.03  & $-$59 16 11.7 & 17.78  & 0.62 & 0.95 & $-$2.34 & 1703 & 4864 \cr
 3 40 29.41  & $-$60 01 29.1 & 16.10  & 0.67 & 0.76 & $-$0.91 & 1125  & 5444
\cr
 3 40 35.41  & $-$59 55 17.5 & 16.25  & 0.59 & 0.65 & $-$0.85 & 1494  & 5848
\cr
 3 40 36.46  & $-$59 15 08.0 & 16.47  & 0.70 & 0.85 & $-$0.60 & 1440  & 5153
\cr
 3 40 38.19  & $-$59 30 54.7 & 16.14  & 0.66 & 0.64 & $-$1.33  & 982 & 5888
\cr
 3 40 40.20  & $-$60 10 09.4 & 16.65  & 0.77 & 0.85 & $-$0.75 & 1234  & 5153
\cr
 3 41 05.77  & $-$59 21 56.2 & 17.48 & 0.67 & 0.79 & $-$1.04 & 2005 & 5343 \cr
 3 41 09.91  & $-$59 09 32.8 & 16.31  & 0.90 & 0.89 & $-$0.19  & 982  & 5033
\cr
 3 41 14.72  & $-$59 26 48.5 & 16.27  & 0.72 & 0.69 & $-$0.23 & 1495  & 5694
\cr
 3 41 14.10  & $-$60 02 07.8 & 17.28  & 0.64 & 0.87 & $-$0.98 & 2011  & 5092
\cr
 3 41 15.39  & $-$59 34 54.9 & 16.26  & 0.87 & 0.91 & $-$0.11 & 1068  & 4975
\cr
 3 41 21.52  & $-$60 06 20.5 & 16.94  & 0.77 & 1.04 & $-$0.77 & 1397  & 4630
\cr
 3 41 22.03  & $-$59 24 33.7 & 17.01  & 0.75 & 0.80 & $-$0.62 & 1621 & 5311 \cr
 3 41 30.33  & $-$60 05 08.8 & 16.48  & 0.76 & 0.85 & $-$0.86 & 1112  & 5153
\cr
 3 41 33.32  & $-$59 13 32.3 & 16.30  & 0.59 & 0.75 & $-$1.91 & 1013 &  5478
\cr
 3 41 33.79  & $-$59 17 43.5 & 17.66  & 0.86 & 1.00 & $-$0.71 & 1615 & 4731 \cr
 3 41 34.72  & $-$60 04 26.7 & 17.48  & 0.63 & 0.80 & $-$0.82 & 2432  & 5311
\cr
 3 41 36.49  & $-$59 48 47.5 & 16.38  & 0.80 & 0.88 & $-$1.93  & 676  & 5063
\cr
 3 41 41.81  & $-$60 04 14.1 & 17.31  & 0.79 & 0.94 & $-$0.71 & 1623 & 4891
\cr
 3 41 45.76  & $-$59 54 08.6 & 16.48  & 0.73 & 0.79 & $-$0.63 & 1326  & 5343
\cr
 \noalign{ \vskip 8pt\hrule height 1pt
  \vskip 8pt }
}}$$
\vfill\eject
  $$ \vbox{ \halign to \hsize { \hfil#\hfil & \hfil#\hfil & \hfil#\hfil &
  \hfil#\hfil & \hfil#\hfil & \hfil#\hfil & \hfil# & \hfil#\hfil \cr
 \multispan{8} \hfil {TABLE 4 (continued)}
\hfil \cr
 \noalign{ \vskip 12pt }
 \noalign{ \hrule height 1pt }
 \noalign{ \vskip 8pt }
      ra (1950.0) &   dec (1950.0)  &    V &   B--V &  V--I & [Fe/H]&
dist(pc)&  Teff
\cr
 \noalign{ \vskip 8pt\hrule height 1pt
  \vskip 8pt }
 3 41 49.99  & $-$60 11 16.4 & 17.50  & 0.90 & 0.97 & $-$0.34 & 1597  & 4810
\cr
 3 41 50.25  & $-$59 22 29.8 & 16.73  & 0.73 & 0.86 & $-$0.63 & 1488  & 5122
\cr
 3 41 52.43  & $-$59 45 18.0 & 16.34  & 0.70 & 0.77 & $-$0.93 & 1162  & 5410
\cr
 3 41 52.49  & $-$59 35 19.7 & 16.12  & 0.66 & 0.73 & $-$0.33 & 1546  & 5549
\cr
 3 41 59.53  & $-$59 49 54.1 & 16.49  & 0.87 & 0.88 & $-$0.54  & 990  & 5063
\cr
 3 42 02.62  & $-$60 06 13.7 & 17.08  & 0.65 & 0.75 & $-$0.68 &
2067  & 5478 \cr
 3 42 13.88  & $-$60 12 33.1 & 17.38  & 0.57 & 0.85 & $-$0.81 & 2686  & 5153
\cr
 3 42 14.96  & $-$59 19 35.2 & 17.07  & 0.73 & 0.77 & $-$0.05 & 2270 & 5410 \cr
 3 42 24.18  & $-$59 12 34.9 & 16.57  & 0.64 & 0.76 & $-$0.84 & 1548  & 5444
\cr
 3 42 37.74  & $-$60 08 41.8 & 17.54  & 0.52 & 0.73 & $-$0.29 & 4310  & 5549
\cr
 3 42 46.91  & $-$60 10 48.8 & 17.40  & 0.81 & 0.90 & $-$0.55 & 1734 & 5004
\cr
 3 42 47.76  & $-$59 46 17.0 & 16.59  & 0.72 & 0.79 & $-$0.19 & 1764  & 5343
\cr
 3 42 59.96  & $-$59 59 13.4 & 17.87  & 0.61 & 0.90 & $-$1.01 & 2788  & 5004
\cr
 3 43 02.04  & $-$59 25 16.7 & 16.14  & 0.59 & 0.71 & $-$0.61 & 1604  & 5620
\cr
 3 43 07.72  & $-$59 12 44.3 & 17.20  & 0.64 & 0.70 & $-$0.42 & 2553 & 5657 \cr
 3 43 07.28  & $-$60  6 21.9 & 16.95  & 0.74 & 0.89 & $-$0.67 & 1578  & 5033
\cr
 3 43 15.04  & $-$59 20 20.2 & 17.46  & 0.79 & 0.76 & $-$0.68 & 1763 & 5444 \cr
 3 43 16.47  & $-$59 43 59.7 & 17.03  & 0.89 & 1.00 & $-$0.85 & 1063  & 4731
\cr
 3 43 23.61  & $-$59 34 27.1 & 17.75  & 0.64 & 0.87 & $-$2.43 & 1588 & 5092 \cr
 3 43 28.02  & $-$59 31 26.8 & 17.13  & 0.65 & 0.74 & $-$0.48 & 2339 & 5513 \cr
 3 43 31.96  & $-$60 15 46.0 & 16.09  & 0.67 & 0.74 & $-$0.57 & 1319  & 5513
\cr
 3 43 33.59  & $-$59 08 54.1 & 17.52  & 0.60 & 0.76  & $-$0.40 & 3300 & 5444
\cr
 3 43 35.01  & $-$59 10 38.8 & 17.58  & 0.81 & 0.97 & $-$0.55 & 1884 & 4810 \cr
 \noalign{ \vskip 8pt\hrule height 1pt
  \vskip 8pt }
}}$$
\vfill\eject
  $$ \vbox{ \halign to \hsize { \hfil#\hfil & \hfil#\hfil & \hfil#\hfil &
  \hfil#\hfil & \hfil#\hfil & \hfil#\hfil & \hfil# & \hfil#\hfil \cr
 \multispan{8} \hfil {TABLE 4 (continued)}
\hfil \cr
 \noalign{ \vskip 12pt }
 \noalign{ \hrule height 1pt }
 \noalign{ \vskip 8pt }
  ra (1950.0) & dec (1950.0)  &  V & B--V &  V--I & [Fe/H]&  dist(pc)&  Teff
\cr
 \noalign{ \vskip 8pt\hrule height 1pt
  \vskip 8pt }
 3 43 37.73  & $-$59 57 07.5 & 16.16  & 0.58 & 0.71 & $-$0.75 & 1542  & 5620
\cr
 3 43 41.02  & $-$59 06 41.0 & 16.18  & 0.57 & 0.68 & $-$0.89 & 1486  & 5732
\cr
 3 43 41.58  & $-$60 05 24.1 & 17.04  & 0.79 & 0.77 & $-$1.58 & 1029  & 5410
\cr
 3 43 47.14  & $-$59 51 31.0 & 16.43  & 0.76 & 0.81 & $-$0.56 & 1246  & 5278
\cr
 3 43 50.94  & $-$59 24 05.5 & 17.04  & 0.75 & 0.73 & $-$0.13 & 2063 & 5549 \cr
 3 43 51.46  & $-$59 50 01.5 & 17.23  & 0.64 & 0.80 & $-$0.29 & 2765  & 5311
\cr
 3 43 56.92  & $-$59 15 15.0 & 17.31  & 0.80 & 0.84  & $-$0.06 &  2110 & 5184
\cr
 3 44 02.10  & $-$59 30 13.6 & 16.27  & 0.68 & 0.73  & $-$0.59 & 1385  & 5549
\cr
 3 44 04.90  & $-$59 24 14.8 & 16.36  & 0.65 & 0.68 & $-$0.11&  1965 &  5732
\cr
 3 44 07.51  & $-$59 14 52.6 & 16.77  & 0.86 & 0.85 & $-$0.27 & 1299  & 5153
\cr
 3 44 11.51  & $-$59 13 45.9 & 16.08  & 0.65 & 0.74 &  $-$0.98 & 1131  & 5513
\cr
 3 44 12.45  & $-$59 52 49.2 & 16.94  & 0.63 & 0.79 & $-$0.54 & 2182  & 5343
\cr
 3 44 13.68  & $-$59 07 36.3 & 16.17  & 0.65 & 0.72 & $-$0.41 & 1558  & 5584
\cr
 3 44 15.06  & $-$60 06 07.3 & 17.28  & 0.77 & 0.80 & $-$0.06 & 2248  & 5311
\cr
 3 44 23.46  & $-$59 55 49.5 & 16.93  & 0.60 & 0.73 & $-$0.55 & 2324  & 5549
\cr
 3 44 29.25  & $-$59 23 07.0 & 16.43  & 0.81 & 0.78 & $-$0.49 &  1140 &  5376
\cr
 3 47 45.83  & $-$60 13 10.3 & 16.33  & 0.66 & 0.78 & $-$1.87 &  897 &  5376
\cr
 3 50 06.66  & $-$60 13 22.6 & 16.11  & 0.78 & 0.78 & $-$0.13 &  1245  & 5376
\cr
 3 50 41.98  & $-$60 32 36.2 & 16.93  & 0.69 & 0.91 & $-$1.35 & 1313 &   4975
\cr
 3 50 46.34  & $-$60 35 14.1 & 16.37  & 0.77 & 0.86 & $-$0.53 & 1200  & 5122
\cr
 3 51 11.32  & $-$60 44 26.3 & 16.37  & 0.69 & 0.80 & $-$0.49 & 1487 &  5311
\cr
 3 51 33.91  & $-$60 45 39.2 & 16.74  & 0.54  & 0.74 & $-$0.38 & 2702  & 5513
\cr
 3 51 37.50  & $-$60 52 26.4 & 16.22  & 0.84  & 0.86 & $-$0.58 &  924 &  5122
\cr
 \noalign{ \vskip 8pt\hrule height 1pt
  \vskip 8pt }
}}$$
\vfill\eject
  $$ \vbox{ \halign to \hsize { \hfil#\hfil & \hfil#\hfil & \hfil#\hfil &
  \hfil#\hfil & \hfil# & \hfil#\hfil & \hfil#\hfil & \hfil#\hfil \cr
 \multispan{8} \hfil {TABLE 4 (continued) }
\hfil \cr
 \noalign{ \vskip 12pt }
 \noalign{ \hrule height 1pt }
 \noalign{ \vskip 8pt }
  ra (1950.0) & dec (1950.0)  &  V & B--V &  V--I & [Fe/H]&  dist(pc)&  Teff
\cr
 \noalign{ \vskip 8pt\hrule height 1pt
  \vskip 8pt }
 3 51 46.54  & $-$60 41 29.1 & 16.18  & 0.78 & 0.70 & $-$0.48 & 1099  & 5657
\cr
 3 51 50.00  & $-$60 24 16.3 & 16.08  & 0.64  & 0.77 & $-$0.06 & 1808  & 5410
\cr
 3 52 39.52  & $-$60 21 08.3 & 16.23  & 0.65 & 0.78 & $-$0.05 & 1895  & 5376
\cr
 3 53 03.53  & $-$60 42 53.1 & 16.85  & 0.74  & 0.74 & $-$0.21 & 1874 &  5513
\cr
 3 53 23.33  & $-$60 46 31.7 & 16.58  & 0.55  & 0.69 & $-$1.19 & 1626  & 5694
\cr
 3 53 42.11  & $-$60 32 10.9 & 16.58  & 0.50 & 0.70 & $-$0.68&  2347 &  5657
\cr
 3 53 45.53  & $-$60 38 08.6 & 16.52  & 0.85 & 0.85 & $-$0.14 & 1253  & 5153
\cr
 3 53 53.94  & $-$60 51 15.9 & 16.57  & 0.73 & 0.72 & $-$0.29 & 1628  & 5584
\cr
 3 53 54.44  & $-$60 35 28.5 & 16.90  & 0.74 & 0.80 & $-$0.36 & 1788  & 5311
\cr
 3 54 40.75  & $-$60 47 58.8 & 16.12  & 0.53 & 0.66 & $-$0.21 & 2279  & 5809
\cr
 3 55 23.73  & $-$60 35 14.1 & 16.07  & 0.67 & 0.72 & $-$0.26 & 1525  & 5584
\cr
 3 59 39.11 & $-$60 11 29.5 & 15.18  & 0.91  & 0.93   & \phantom{$-$}0.30     &
613      & 4919 \cr
 3 59 59.63 & $-$60 15 05.5 & 15.23 &  0.69 & 0.70 & \phantom{$-$}0.32   &
1111  & 5657 \cr
 3 59 02.73 & $-$60 14 49.7 & 15.06 & 0.60 & 0.67 &  $-$0.69
& 914     & 5770 \cr
3 59 52.51 & $-$60 18 10.7 & 15.90 & 0.77 & 0.87 & $-$0.61 & 931 & 5092 \cr
3 59 38.20 & $-$60 19 19.6 & 15.38 & 0.57 & 0.70 &  $-$0.67 & 1149 & 5657 \cr
3 59 01.68 & $-$60 23 40.1 & 16.05 & 0.92  & 0.95 & $-$0.54  & 716  & 4864 \cr
3 58 13.33 & $-$60 19 11.7 & 15.00    &  0.68  & 0.77  & \phantom{$-$}0.26   &
1026      & 5410 \cr
 \noalign{ \vskip 8pt\hrule height 1pt \vskip 1pt \hrule height 1pt
  \vskip 8pt }
}}$$

\vfill\eject

\parskip=3pt plus 1pt minus 1pt  %it is zero plus 1pt in plain.tex
  \tabskip=.2em plus .5em minus .2em
  \def \space{ \noalign{ \vskip4pt \hrule height 1pt \vskip4pt} }
  $$ \vbox{ \halign to \hsize { \hfil#\hfil & \hfil#  & \hfil#  &
  \hfil#  & \hfil# & \hfil#  \cr
 \multispan{6} \hfil {TABLE 5 : {F117  Iron Abundance Distribution }}\hfil \cr
 \noalign{ \vskip 12pt }
 \noalign{ \hrule height 1pt \vskip 1pt \hrule height 1pt }
 \noalign{ \vskip 8pt }
Central [Fe/H] & 1  & 2 & 3 & 4 & 5 \cr
 \noalign{ \vskip 8pt \hrule height 1pt }
 \noalign{ \vskip 8pt }
+0.3 & 2 &  2350 & 2035 & 2 & 6 \cr
+0.1 & 0  & 2350 & 2035 & 0 & 0 \cr
$-$0.1 &  11 & 2240 & 2210 &  12 & 13 \cr
$-$0.3 &  13 & 2045 & 2015 & 16 & 18 \cr
$-$0.5 &  17 & 1850 & 1850 & 26 & 26 \cr
$-$0.7 & 17 &  1670 & 1670 & 32 & 32 \cr
$-$0.9 & 12 &  1520 & 1520 & 28 & 28 \cr
$-$1.1 & 3 & 1390 & 1430 & 8 & 9 \cr
$-$1.3 & 2 & 1280 & 1315 & 7 & 7 \cr
$-$1.5 & 1 & 1190 & 1215 & 4 & 5 \cr
$-$1.7 & 0 & 1110  & 1135 & 0 & 0 \cr
$-$1.9 & 3 & 1045 & 1070 & 17 & 20 \cr
$-$2.1 & 0 &  990 & 1015 & 0 & 0  \cr
$-$2.3 & 1 & 950 & 970 & 7 & 9 \cr
$-$2.5 & 1 & 915 &   935 & 8 & 11 \cr
 \noalign{ \vskip 8pt\hrule height 1pt \vskip 1pt \hrule height 1pt
  \vskip 8pt }
}}$$
\baselineskip=12pt

\centerline {Notes to TABLE 5}

 Column number 1 contains the `raw' observed metallicity
distribution for the  stars with 0.5$\le {\rm B-V} < 0.9$; column
number 2
the characteristic distance (in pc) of each metallicity bin if all stars are
assumed to follow the thick disk density profile; column number 4
the number of stars per unit volume at a fiducial vertical height
above the plane
distance of 1000pc under this assumed
density law; column number 3 the characteristic distances if
stellar halo and thin disk contributions are introduced, as explained
in the text; column number 5  the number of stars
per unit volume at $ z = 1000$pc under these assumed density laws.
\vfill\eject
\baselineskip=20pt
\parskip=3pt plus 1pt minus 1pt  %it is zero plus 1pt in plain.tex
  \tabskip=.2em plus .5em minus .2em
  \def \space{ \noalign{ \vskip4pt \hrule height 1pt \vskip4pt} }
  $$ \vbox{ \halign to \hsize { \hfil#\hfil & \hfil#  & \hfil#  \cr
 \multispan{3} \hfil {TABLE 6 : {Combined   Iron Abundance
Distribution at $Z=1500$pc}}\hfil \cr
 \noalign{ \vskip 12pt }
 \noalign{ \hrule height 1pt \vskip 1pt \hrule height 1pt }
 \noalign{ \vskip 8pt }
Central [Fe/H] & All Thick Disk  & 3 Components  \cr
 \noalign{ \vskip 8pt \hrule height 1pt }
 \noalign{ \vskip 8pt }
+0.3 & 3 &  7 \cr
+0.1 & 8  & 22 \cr
$-$0.1 &  22 & 24 \cr
$-$0.3 &  38 & 39 \cr
$-$0.5 &  45 & 45 \cr
$-$0.7 & 46 & 46 \cr
$-$0.9 & 45 & 45 \cr
$-$1.1 & 13 & 17 \cr
$-$1.3 & 16 & 20 \cr
$-$1.5 & 16 & 23 \cr
$-$1.7 & 0 & 0 \cr
$-$1.9 & 15 & 25 \cr
$-$2.1 & 0 &  0 \cr
$-$2.3 & 4 & 8 \cr
$-$2.5 & 17 & 29 \cr
 \noalign{ \vskip 8pt\hrule height 1pt \vskip 1pt \hrule height 1pt
  \vskip 8pt }
}}$$

\bye